\documentclass[twocolumn,english,showpacs,footinbib,bibnotes]{revtex4-1}

\usepackage[version=3]{mhchem}

\usepackage{graphicx}
\usepackage[usenames,dvipsnames]{color}
\usepackage{dcolumn}
\usepackage{bbm}
\usepackage{bm}
\usepackage{amsmath}
\usepackage{amssymb}
\usepackage{times}
\usepackage{placeins}
\usepackage{hyperref}
\hypersetup{colorlinks=false}

\begin{document}

\title{Competing magnetic orders and spin liquids in two- and three-dimensional kagome systems: \\ 
	A pseudo-fermion functional renormalization group perspective}

\author{Finn Lasse Buessen}
\author{Simon Trebst}
\affiliation{Institute for Theoretical Physics, University of Cologne, 50937 Cologne, Germany}

\date{\today}

\begin{abstract}
Quantum magnets on kagome lattice geometries in two and three spatial dimensions are archetypal examples of spin systems
in which geometric frustration inhibits conventional magnetic ordering and instead benefits the emergence of long-range entangled 
spin liquids at low temperature. Here we employ a recently developed pseudo-fermion functional renormalization
group (pf-FRG) approach to study the low-temperature quantum magnetism of kagome and hyperkagome spin systems 
with exchange interactions beyond the nearest neighbor coupling. We find that next-nearest neighbor couplings stabilize
a variety of magnetic orders as well as induce additional spin liquid regimes giving rise to rather rich phase diagrams, which we 
characterize in detail. On a technical level, we find that the pf-FRG approach is in excellent quantitative agreement with high-temperature 
series expansions over their range of validity and it exhibits a systematic finite-size convergence in the temperature regime below. We discuss notable advantages and some current limitations of the pf-FRG approach in the ongoing search for unconventional forms of quantum magnetism. 
\end{abstract}

\maketitle


\section{Introduction}
In quantum magnetism, the Heisenberg antiferromagnet on the kagome lattice is one of the most widely studied systems as it is one of the most elementary two-dimensional systems that evades a conventionally ordered ground state \cite{Diep2004}. Geometric frustration arises from the lattice structure of corner-sharing triangles,  
it inhibits the formation of any local order, and leads to the formation of long-range entanglement -- the signature of so-called quantum spin liquids \cite{Savary2016,Balents2010,Imai2016}. However, the precise nature of this spin liquid ground state has remained under debate for decades including proposals for topologically ordered $Z_2$ spin liquids \cite{Read1991,Wen1991,Sachdev1992,Misguich2002,Wang2006,Lu2011,Depenbrock2012,Jiang2012,Mei2016} with an extremely small energy gap \cite{Yan2011,Laeuchli2016} as well as algebraic $U(1)$ Dirac spin liquids \cite{Hastings2000,Ran2007,Hermele2008,Iqbal2013,He2016}, which are gapless and only marginally stable \cite{Hermele2004}. Theoretical interest in kagome systems has recently been further spurred by the unambiguous numerical observation \cite{Bauer2014,He2014,Gong2014,Gong2015} of a chiral spin liquid ground state \cite{Kalmeyer1987} -- a long sought-after bosonic analogue of the fractional quantum Hall effect -- in kagome systems augmented by next-nearest neighbor interactions or chiral interactions. Experimental interest in kagome spin liquids \cite{Mendels2016} has been driven by the synthesis of herbertsmithite ZnCu$_3$(OH)$_6$Cl$_2$ \cite{Herbertsmithite}, the hitherto cleanest material realization of the kagome lattice structure retaining its three-fold lattice symmetry and having a dominant nearest-neighbor antiferromagnetic Heisenberg exchange \cite{Norman2016}, as well as the copper minerals volborthite, Cu$_3$V$_2$O$_7$(OH)$_2\cdot$2H$_2$O \cite{Volborthite}, and kapellasite, Cu$_3$Zn(OH)$_6$Cl$_2$ \cite{Kapellasite}, which both exhibit substantial magnetic exchange couplings beyond the nearest neighbor interaction.

\begin{figure}[b]
  \centering
  \includegraphics[width=\linewidth]{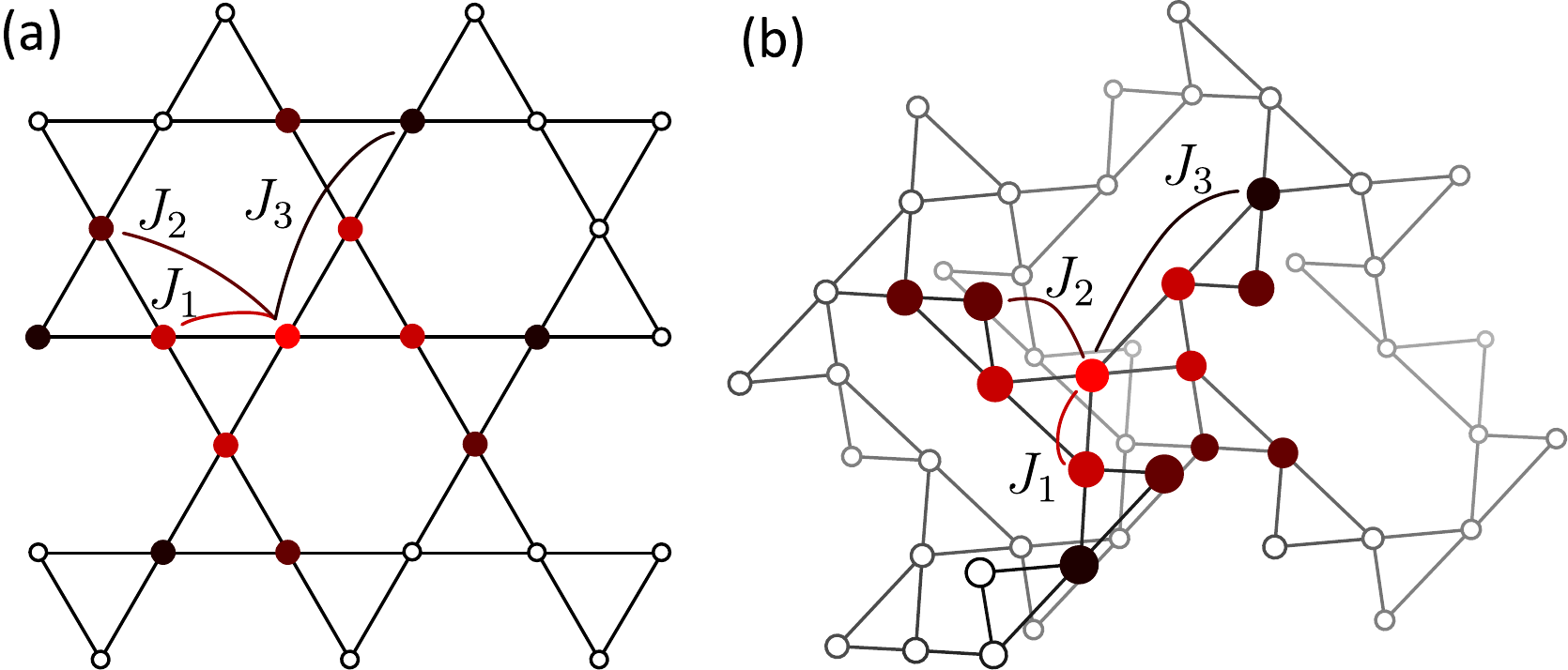}
  \caption{ (a) Two-dimensional kagome lattice with nearest neighbor couplings  $J_1$, next-nearest neighbor couplings $J_2$ and a subset of 
  				third-nearest neighbors coupled with strength $J_3$.
		(b) Three-dimensional hyperkagome lattice, which like the kagome lattice is a network of corner-sharing triangles.
		     	Exchange couplings between nearest neighbors are parametrized by $J_1$. Next-nearest and third-nearest neighbors at bond distance 2
			are parametrized by $J_2$ and $J_3$, respectively.
  		       Note that the kagome lattice has four next-nearest neighbors coupled via $J_2$ and four third-nearest neighbors coupled via $J_3$, respectively, 
		       while the hyperkagome lattice has six next-nearest neighbors coupled via $J_2$ and only two third-nearest neighbors coupled via $J_3$. }
  \label{fig:lattices}
\end{figure}

The three-dimensional cousin of the kagome lattice is the so-called hyperkagome lattice \cite{Okamoto2007} -- a three-dimensional network of corner-sharing triangles
illustrated in Fig.~\ref{fig:lattices}(b). The hyperkagome lattice is a body-centered cubic lattice that is sometimes also referred to as half-garnet lattice, which emphasizes the fact that the lattice is lacking inversion symmetry. Its geometric nature can be understood by considering the premedial lattice spanned by the elementary triangles -- while for the two-dimensional kagome lattice this premedial lattice is the well-known honeycomb lattice, the premedial lattice of the hyperkagome is the so-called hyperoctagon \cite{Hermanns2014} or (10,3)a lattice \cite{Wells1977}, one of the most elementary tricoordinated lattices in three spatial dimensions which has recently attracted some interest in the context of three-dimensional Kitaev models \cite{OBrien2016}. 
The quantum magnetism of the hyperkagome lattice has attracted considerable theoretical interest
\cite{Hopkinson2007,Lawler2008a,Zhou2008,Lawler2008b,Chen2008,Podolsky2009,Motome2009,Norman2010,Micklitz2010,Bergholtz2010,Kim2011,Singh2012,Chen2013,Kimchi2014,Shindou2016,Kim2016,Kim2016b} since the synthesis of the iridate compound Na$_4$Ir$_3$O$_8$  \cite{Okamoto2007}, which to date remains the best candidate material for a three-dimensional spin liquid compound. However, very much like for its two-dimensional counterpart, the determination of the precise nature of this hyperkagome spin liquid has remained challenging. For the pure Heisenberg antiferromagnet on the hyperkagome lattice, theoretical proposals include the emergence of a gapped $Z_2$ spin liquid with topological order \cite{Lawler2008a}, a gapless $U(1)$ spin liquid with a spinon Fermi surface \cite{Zhou2008,Lawler2008b}, and the formation of a valence bond crystal \cite{Bergholtz2010}. In addition, several groups have argued to augment the nearest-neighbor Heisenberg exchange by the inclusion of spin-orbit coupling effects \cite{Chen2008,Podolsky2009,Norman2010,Micklitz2010}, e.g. by considering a Dzyaloshinskii-Moriya interaction, a Kitaev-like exchange term or additional (symmetric) anisotropic couplings \cite{Shindou2016,Kim2016}.

In this manuscript, we study the formation of spin liquids and magnetic order in kagome and hyperkagome systems by considering a generalized SU(2) spin-1/2 Heisenberg model augmented with exchange terms beyond the bare nearest neighbor exchange
\begin{equation}
	H=J_1\sum_{\langle i,j\rangle}{\bf S}_i {\bf S}_j + J_2\sum_{\langle\langle i,j\rangle\rangle_\wedge}{\bf S}_i {\bf S}_j + J_3\sum_{\langle\langle i,j\rangle\rangle_{\textemdash}}{\bf S}_i {\bf S}_j,
\label{eq:j1j2j2bHamiltonian}
\end{equation}
where the spin operators ${\bf S}_i$ describe quantum mechanical spin-1/2 moments. The first sum $\langle i,j \rangle$ runs
over pairs of nearest neighbors $i$ and $j$. The second and third sums run over next-nearest and third-nearest neighbors, which are at bond distance 2 but slightly differ in their bond geometry. The sum over $\langle\langle i,j\rangle\rangle_\wedge$ runs over pairs of next-nearest neighbors coupled via two {\em angled} bonds, while the third sum over $\langle\langle i,j\rangle\rangle_{\textemdash}$ runs over third-nearest neighbors coupled via two {\em collinear} bonds, see Fig.~\ref{fig:lattices}.
Note that such a definition of next-nearest and third-nearest neighbors via their bond distance is not identical to a definition in terms of spatial distance.
For the kagome lattice, we do not include a coupling diagonally across the hexagon in our $J_3$ couplings (although they are at the same spatial distance), while for the hyperkagome lattice we similarly exclude additional couplings across the elementary decagon in our $J_2$ and $J_3$ couplings.
Microscopically, the dominant exchange path for such couplings across the elementary hexagon or decagon plaquettes arises via a magnetically neutral atom in the center of the plaquettes, which however is absent for the kagome material herbertsmithite and the hypergakome material Na$_4$Ir$_3$O$_8$. 
In the following, we consider exchange constants $J_1, J_2$ and $J_3$ that can be either antiferromagnetic (positive) or ferromagnetic (negative). We start our discussion with the pure Heisenberg antiferromagnet ($J_1 = 1, J_2 = J_3 = 0$), then consider the effect of the angled coupling $J_2$ and finally include the collinear bond coupling $J_3$.

To explore the physics of the generalized Heisenberg model \eqref{eq:j1j2j2bHamiltonian}, we apply the pseudo-fermion functional renormalization group (pf-FRG)  recently developed by Reuther and W\"olfle \cite{Reuther2010}. The key idea is to express the spin degrees of freedom in terms of auxiliary complex fermions (or pseudo\-fermions)
and to subsequently apply a functional renormalization group approach \cite{Wetterich1993}, which has been extensively used in the context of interacting many-fermion systems  \cite{Kopietz2010}. This procedure has proven to be a remarkably versatile and accurate tool to detect magnetic ordering in frustrated quantum magnets and in some instances to also provide evidence for potential spin liquid phases \cite{Reuther2010,Reuther2011,Reuther2011a,Reuther2011b,Reuther2011c,Reuther2012,Reuther2014,Reuther2014a,Rousochatzakis2015,Suttner2014,Iqbal2015,Iqbal2016,Iqbal2016b}. The pf-FRG approach thereby naturally complements the well-established toolbox of numerical methods, which includes quantum Monte Carlo approaches (typically plagued by a severe sign problem  for frustrated quantum magnets), exact diagonalization techniques (limited by finite system sizes), the density matrix renormalization group (for moderately entangled systems), and series expansion techniques (requiring sophisticated series extrapolation techniques). Similar to the series expansion techniques the pf-FRG approach can handle competing interactions (without encountering a sign problem) and is per se not limited to one or two spatial dimensions. Applying the pf-FRG method to three-dimensional systems requires a careful implementation of lattice symmetries but then allows to consider systems of a few hundred sites, as we will discuss in more detail in the next Section. A first step to explore the physics of three-dimensional frustrated quantum magnets with the pf-FRG approach has recently been taken for a frustrated $J_1J_2J_3$ antiferromagnet on the highly symmetric cubic lattice \cite{Iqbal2016}.

\begin{figure}
  \centering
  \includegraphics[width=\linewidth]{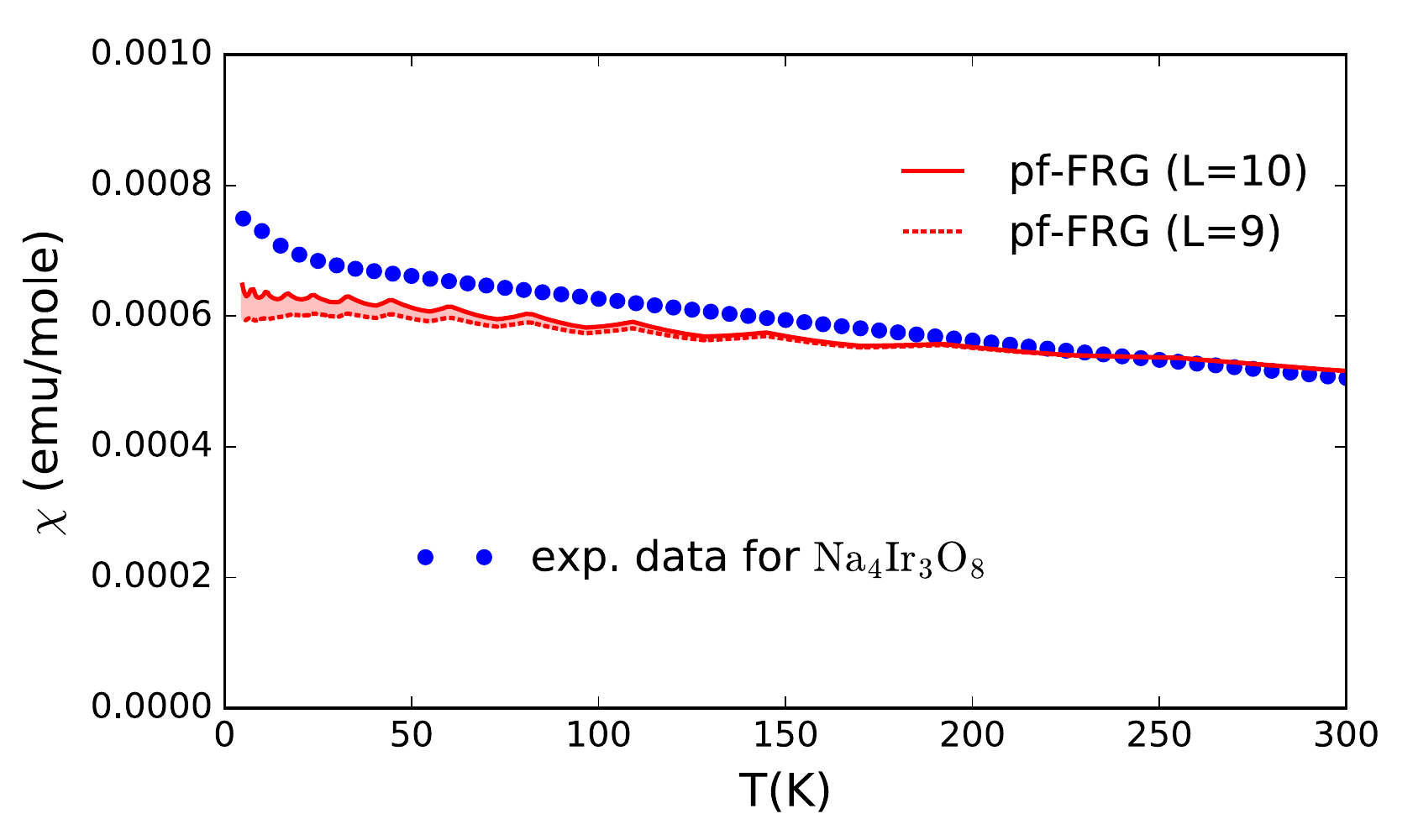}
  \caption{Uniform susceptibility of the Heisenberg antiferromagnet on the hyperkagome lattice obtained from pf-FRG calculations
  		in comparison with experimental data obtained for \ce{Na4Ir3O8} \cite{Okamoto2007}.
		}
  \label{fig:hyperkagomeAFMSusceptibilityExperiment}
\end{figure}

We apply the pf-FRG approach to the kagome and hyperkagome systems introduced above. Our focus is on the quantum magnetism induced by next-nearest and third-nearest neighbor exchanges and we provide a number of phase diagrams with a zoo of magnetically ordered states and potential spin liquid phases in the remainder of the manuscript. Here we highlight one principal result of our pf-FRG calculations obtained for the pure Heisenberg antiferromagnet on the hyperkagome lattice. Shown in Fig.~\ref{fig:hyperkagomeAFMSusceptibilityExperiment} is the magnetic susceptibility $\chi$ obtained from pf-FRG calculations for the two largest system sizes accessible in our calculations with 573 sites ($L=9$) and 785 sites ($L=10$), respectively. The pf-FRG data, which shows almost no finite-size dependence down to zero temperature, exhibits no divergence of $\chi$ indicative of the absence of a magnetic ordering transition and is thus in line with a possible spin liquid ground state \cite{FootnoteSpinLiquidCriterion}. Even more interesting is the rather comprehensive agreement with the experimental susceptibility obtained for Na$_4$Ir$_3$O$_8$, which shows a moderate deviation from the pf-FRG data only below 150~K and a slight upturn not present in the pf-FRG data around 20~K. We take this as a strong indicator that the physics of Na$_4$Ir$_3$O$_8$ is rather close to the spin liquid ground state of the pure Heisenberg antiferromagnet. More generally, this result  already attests to the usefulness of the pf-FRG in complementing other numerical and analytical approaches. 

Our discussion in the following is organized as follows. We start out in Sec.~\ref{sec:pfFRG} with a short review of the pseudo\-fermion functional renormalization group approach including a detailed description of its efficient adaptation to three-dimensional lattice structures and a comparison to high-temperature series expansions. We then turn to the quantum magnetism of the generalized Heisenberg model \eqref{eq:j1j2j2bHamiltonian} on the hyperkagome lattice in Sec.~\ref{sec:hyperkagome}, in which we provide a detailed account of the magnetic ordering and spin liquids induced by the next-nearest and third-nearest neighbor coupling. In Sec.~\ref{sec:kagome} we turn our discussion  to the kagome system and close with a short summary of our results and an outlook in Sec.~\ref{sec:discussion}.


\section{Pseudo-fermion functional renormalization group}
\label{sec:pfFRG}

Before we turn to the adaptation of the pf-FRG method to the two- and three-dimensional kagome systems in the focus of this manuscript, we start with a short review of the pf-FRG technique \cite{Reuther2010}. The key idea of this approach is to recast all spin operators in terms of auxiliary complex fermions 
\begin{equation}
  S^\mu_i=\frac{1}{2} \sum_{\alpha,\beta} f^\dagger_{i\alpha}\sigma^\mu_{\alpha\beta}f^{\phantom\dagger}_{i\beta} 
  \label{eq:pseudofermions}
\end{equation}
and to then harness the well-developed FRG framework \cite{Wetterich1993} for interacting many-fermion systems  \cite{Kopietz2010}. Note, however, that recasting the magnetic interactions in terms of these auxiliary fermions one naturally obtains only {\it quartic} interaction terms with the  Heisenberg couplings of Eq.~\eqref{eq:j1j2j2bHamiltonian}, for instance, taking the form
\begin{equation}
	{\bf S}_i {\bf S}_j = \frac{1}{4} \sum_{\alpha,\beta,\gamma,\delta,\mu} f^{\dagger}_{i\alpha} f^{\dagger}_{j\gamma} \sigma^{\mu}_{\alpha\beta} \sigma^{\mu}_{\gamma\delta} f^{\phantom\dagger}_{j\delta} f^{\phantom\dagger}_{i\beta} \,.
	\label{eq:pseudofermionInteraction}
\end{equation}
As a consequence, the auxiliary fermion system is somewhat atypical in that it does not comprise any kinetic terms typically present in ordinary many-fermion models. The auxiliary fermion system is thus always in the strong coupling regime defined by the quartic interactions and as such not amenable to conventional perturbation theory. The pf-FRG approach overcomes this lack of a systematic expansion parameter by re-summing an infinite number of diagrams that go beyond ordinary ladder approximations \cite{Kopietz2010}. To this end -- in analogy to ordinary fermionic FRG -- the auxiliary fermion system is treated in the functional integral formalism with an action of the form
\begin{equation}
S=-\int\limits_{1',1}\bar{\psi}_{1'}G_0^{-1}\psi_1 + \int\limits_{1',2',2,1}\bar{\psi}_{1'}\bar{\psi}_{2'}V_{1',2',2,1}\psi_2\psi_1 \,,
\end{equation}
where each collective index should be understood as an integration over frequency, lattice site, and spin index and the fermionic operators have been transformed into Grassmann fields $\bar{\psi}$ and $\psi$. Lacking a kinetic term, the only contribution to the Gaussian propagator is the Matsubara frequency that stems from the functional integral, hence
\begin{equation}
G_0=\frac{1}{i\omega} \,.
\end{equation}
The quartic interaction $V_{1',2',2,1}$ can comprise spin interactions between arbitrary lattice sites according to Eq.~\eqref{eq:pseudofermionInteraction}. 

A renormalization group flow is then generated by introducing a sharp frequency cutoff $\Lambda$ in the Gaussian propagator
\begin{equation}
G_0^\Lambda(\omega)=\frac{\Theta\left(\left|\omega\right|-\Lambda\right)}{i\omega} 
\end{equation}
such that the Gaussian propagator remains unchanged for $\Lambda=0$ and vanishes in the limit of infinite cutoff $\Lambda\to\infty$. The flow equations that connect the particularly simple case of the vanishing propagator to the physically relevant scenario of zero cutoff are conveniently formulated in terms of an effective action and single-particle irreducible (1PI) interaction vertices. In the 1PI formulation the interaction vertices in the limit of infinite cutoff reduce to
\begin{equation*}
\includegraphics[width=0.8\linewidth]{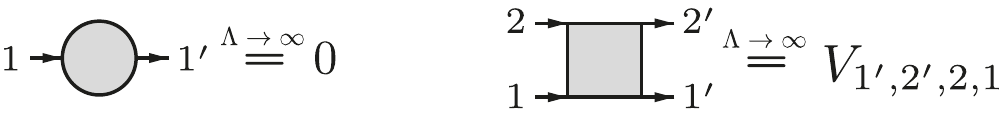} 
\end{equation*}
meaning that all spin interactions are encoded in the initial conditions of the flow at infinite cutoff. All higher order vertices vanish. In principle these flow equations, which can be formally derived by taking the cutoff-derivative of the effective action, provide an analytically exact description of the system, since they can simply be considered to be a formal transformation from an integral formulation to a differential notation. This implies that the high complexity of the system should remain unaltered, which is indeed reflected by the mathematical structure of the flow equations. In general, the flow equation for the $n$-particle interaction vertex may depend on all vertices up to order $n+1$ which means that the resulting system of coupled integro-differential equations never closes. Therefore, for any practical computation, the hierarchy needs to be truncated at a given order. In pf-FRG calculations one typcially includes the 1-particle vertex and the 2-particle vertex for which the flow equations read
\begin{equation*}
\includegraphics[width=0.4\linewidth]{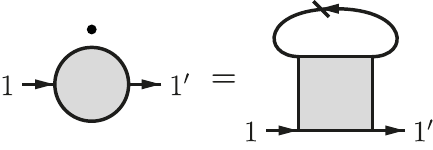}
\end{equation*}
\begin{equation*}
\includegraphics[width=\linewidth]{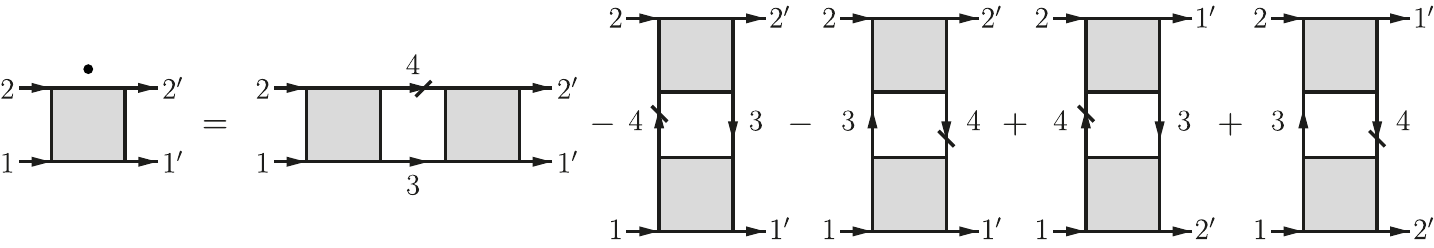} \,,
\end{equation*}
where the arrow denotes, as usual, the full (cutoff-dependent) propagator
\begin{equation}
G^\Lambda(\omega)=\frac{\Theta\left(\left|\omega\right|-\Lambda\right)}{i\omega-\Sigma^\Lambda(\omega)}
\end{equation}
and the slashed arrow denotes the single-scale propagator
\begin{equation}
S^\Lambda(\omega)=\frac{\delta\left(\left|\omega\right|-\Lambda\right)}{i\omega-\Sigma^\Lambda(\omega)} \,.
\end{equation}
Note that the full flow equation for the 2-particle vertex has an additional term that depends on the 3-particle vertex and is zero in this truncation. While it is in principle possible to successively increase the number of vertices that are included in the calculation, already the calculation of the full three-particle vertex turns out to be beyond the computationally feasible. Yet, parts of the three-particle vertex contribution can be recovered by implementing the so-called Katanin scheme \cite{Katanin2004}, which replaces the single-scale propagator by the cutoff derivative of the full propagator
\begin{equation}
S^\Lambda(\omega) \longrightarrow -\frac{d}{d\Lambda}G^\Lambda(\omega)=S^\Lambda(\omega)-\left(G^\Lambda(\omega)\right)^2 \frac{d}{d\Lambda}\Sigma^\Lambda(\omega) \,.
\end{equation}
This generates an additional term that would naturally appear in the 3-particle contribution to the flow of the 2-particle vertex. The Katanin scheme has been found \cite{Reuther2010} to be crucial in order to retrieve sensible results with the pf-FRG approach.

To solve the flow equations numerically we use a logarithmic frequency mesh with 66 points in the range from $-250$ to $250$. We treat systems that are in principle infinitely large but the spin correlations are limited in spatial extent with the system size $L$ being defined as the maximum distance in terms of the number of connecting bonds that two sites with non-zero correlations may be apart. For complex three-dimensional lattice structures, such as the hyperkagome system at hand, typical system sizes that can be efficiently treated are a few hundred lattice sites. In Table~\ref{table:systemSizes} we list the number of differential equations as a function of system size $L$ for the hyperkagome lattice that need to be solved. Since one needs to calculate one flow equation for every possible index combination, one encounters a vast and quickly growing number of coupled differential equations. To reduce the computational effort one can make use of the various symmetries of the problem. This includes the SU(2) spin symmetry of the Heisenberg exchange, which can be captured in a specific ansatz for the interaction vertices \cite{Reuther2010}. In addition, there is a multitude of lattice symmetries that can be identified for a given finite system size $L$. Carefully distilling and implementing these symmetries allows to greatly reduce the number of differential equations, see the second and third column of Table~\ref{table:systemSizes}, and thereby speed up the numerical calculations significantly.

\begin{table}
\begin{tabular}{|c|c|c|c|}
\hline
$L$ & \# sites & \# differential equations & \# differential equations   \\
& & w/o lattice symmetries & with lattice symmetries \\
\hline
5 & 115 & 489,668,883 & 2,147,541 \\
\hline
6 & 185 & 1,267,214,883 & 3,443,451 \\
\hline
7 & 287 & 3,049,794,627 & 5,368,803 \\
\hline
8 & 415 & 6,376,802,883 & 7,738,467 \\
\hline
9 & 573 & 12,156,709,587 & 10,663,521 \\
\hline
10 & 785 & 22,816,346,883 & 14,588,277 \\
\hline
\end{tabular}
\caption{Typical system sizes for the hyperkagome lattice in relation to the number of flow equations that need to be solved either with or without the use of lattice symmetries. }
\label{table:systemSizes}
\end{table}

\begin{figure}[b]
  \centering
  \includegraphics[width=\linewidth]{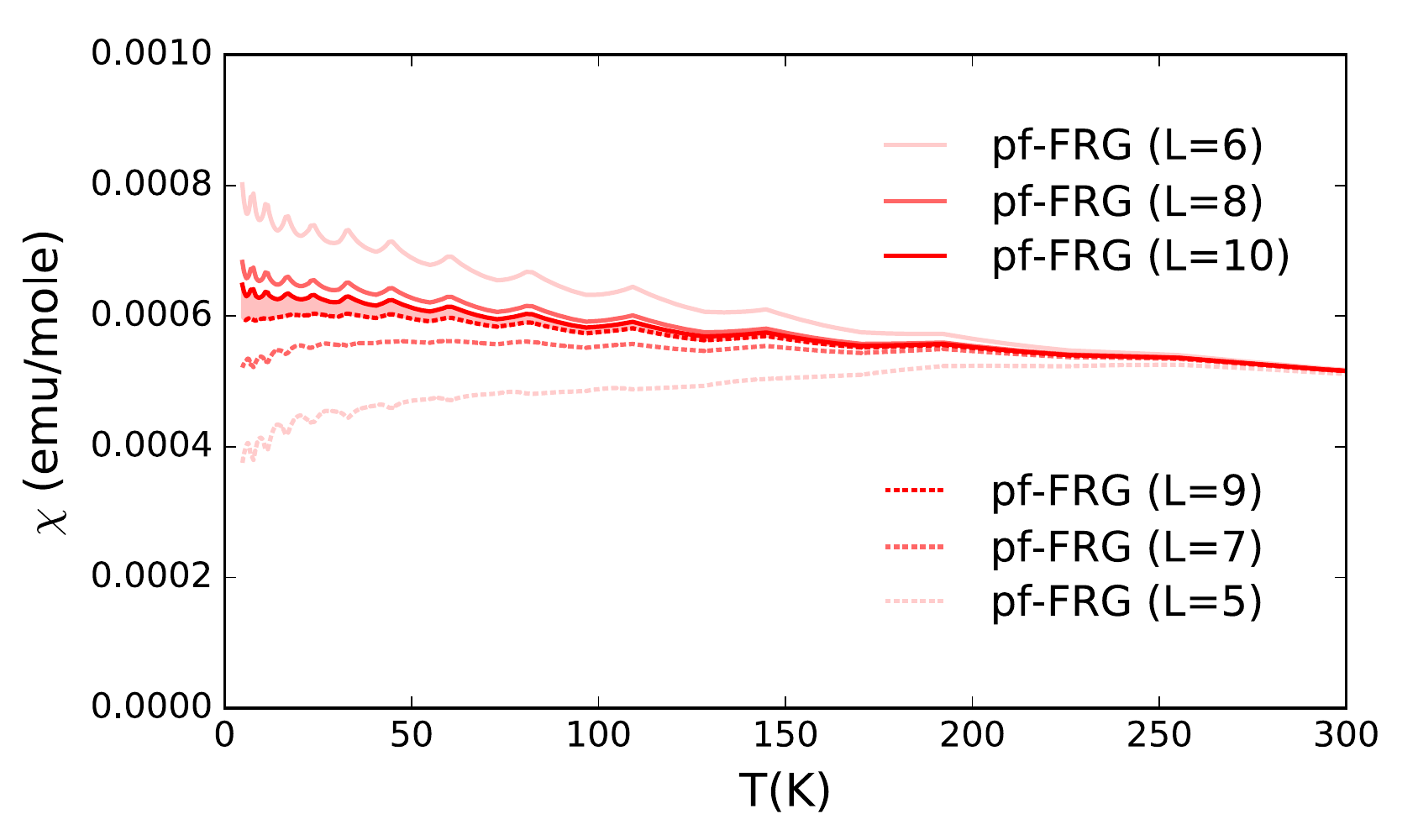}
  \includegraphics[width=\linewidth]{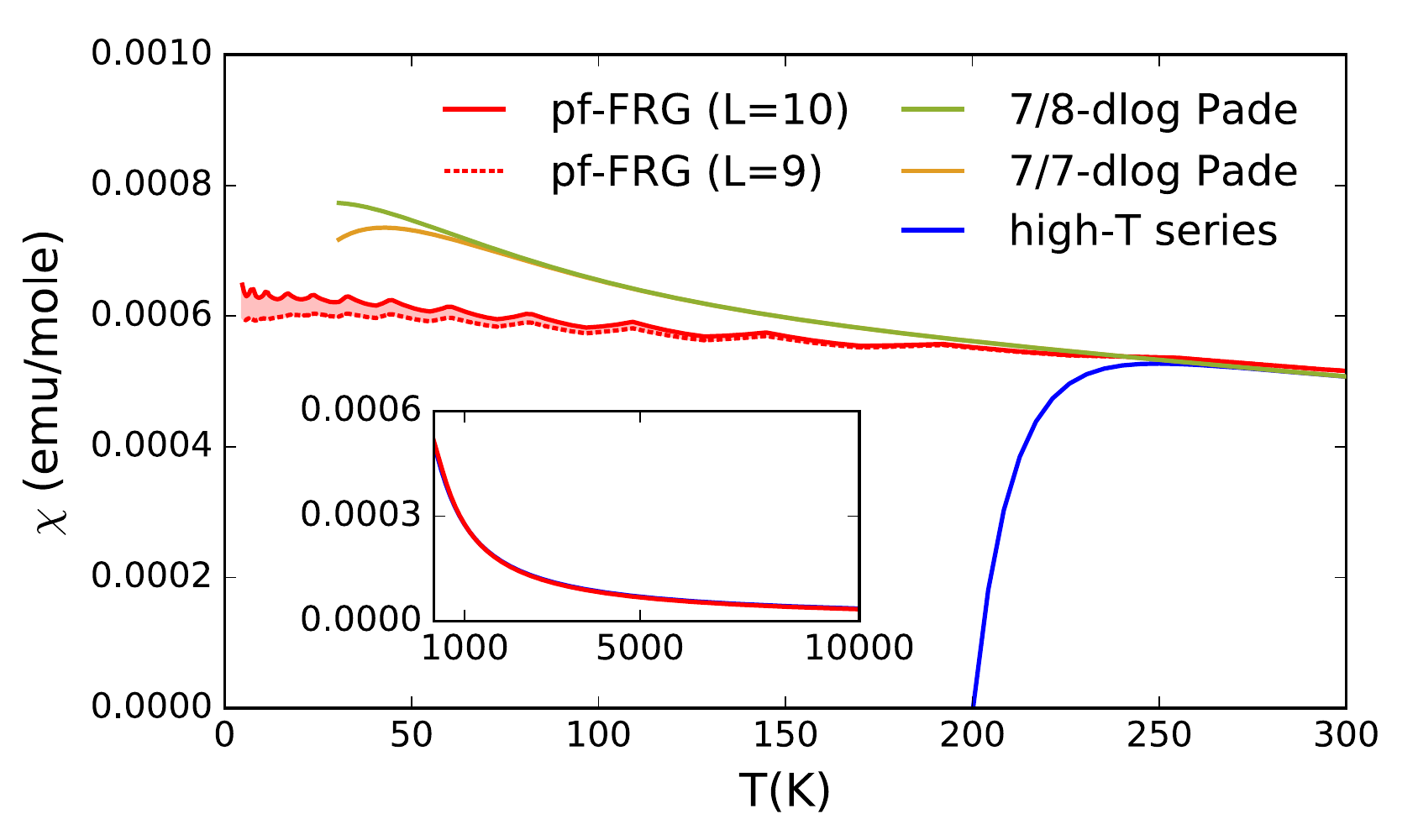}
  \caption{ Top panel: Uniform susceptibility of a Heisenberg AFM on the hyperkagome lattice. 
  		Shown is the convergence of pf-FRG calculations with system size $L$. 
		Systems of even length are found to approach the thermodynamic limit from above, odd system sizes from below,
		leaving a well-defined sliver of finite-size uncertainty between the two largest system sizes ($L=9$ and $L=10$)
		indicated by the opaque red shading.
  		Bottom panel: Comparison with high-temperature series expansion results obtained from a state-of-the-art 
		16-th order expansion in the inverse temperature \cite{Singh2012}. 
		To go beyond the convergence regime of the bare series (blue line), high-order differential Dlog Pad\'e approximants
		to the series are provided (green and orange lines). 
		The inset shows a comparison of the pf-FRG and high-temperature expansion results over a wide temperature range
		with no discernible difference between the two approaches.
  		}
  \label{fig:hyperkagomeAFMSusceptibility}
\end{figure}

To demonstrate the quantitative merits of the pf-FRG approach we first turn to the pure Heisenberg antiferromagnet on the hyperkagome lattice, i.e. model \eqref{eq:j1j2j2bHamiltonian} with $J_1 = 1, J_2 = J_3 = 0$. Probably the most natural observable to identify possible magnetic order is the uniform magnetic susceptibility defined as 
\begin{equation}
\chi=\beta\sum\limits_{i,j}\langle S^z_i S^z_j \rangle 
\end{equation}
with the sum running over all lattice sites $i$ and $j$, and $\beta$ the inverse temperature.
In the pf-FRG formalism this susceptibility can be calculated as an expectation value that is quartic in the fermionic operators and can be expanded diagrammatically in a straight-forward manner \cite{Reuther2010}. We recover physical units by multiplying each spin operator with a magnetic moment of $g\mu_B/2$ where $g=1.99$ is the $g$-factor of the spin moment, $\mu_B$ is the Bohr magneton and we set $J_1=300K$ \cite{Singh2012}. We further identify the frequency cutoff $\Lambda$ with a temperature by noting that the frequency cutoff marks the smallest Matsubara frequency scale in the system and hence should indeed set a temperature scale. Indeed we can think of the re-integration of the flow equations as being in a one-to-one correspondence with the cooling process from infinite temperature down to a given temperature regime -- a perspective that has been used in previous pf-FRG calculations of thermodynamic quantities \cite{Reuther2011c}. A more rigorous treatment \cite{Iqbal2016} reveals the connection between the frequency cutoff $\Lambda$ and the temperature to be precisely given by
\begin{equation}
	\Lambda=2T/\pi \,.
	\label{eq:LambdaT} 
\end{equation} 


Results for the so-rescaled uniform susceptibility obtained from pf-FRG calculations for different system sizes are shown in the 
top panel of Fig.~\ref{fig:hyperkagomeAFMSusceptibility}. We find a very systematic finite-size convergence, with data for even system sizes approaching the thermodynamic limit of infinite system sizes from above, while data for odd system size approaches it from below. This leaves a well-defined sliver of finite-size uncertainty that is found to systematically decrease with increasing system sizes, thus providing us with an intrinsic estimate for the uniform susceptibility including its variation. To quantitatively compare these results we turn to high-temperature series expansions, which in a state-of-the-art calculation for the hyperkagome system have recently been determined up to order 16 in the inverse temperature \cite{Singh2012}. A comparison with our pf-FRG results is provided in the lower panel of Fig.~\ref{fig:hyperkagomeAFMSusceptibility}. The bare series is found to diverge around 250~K, which can be partially lifted by an extrapolation of the series via high-order differential Dlog Pade approximants as illustrated in the figure. Comparing the pf-FRG results to those of the series expansion we find excellent quantitative agreement down to about 180~K, i.e. over the entire range of validity of the series expansion results. This provides us with a crucial quantitative assurance of the validity of the pf-FRG approach for the low-temperature (spin liquid) regime of frustrated quantum magnets, which has been missing hitherto.

\begin{figure}
  \centering
  \includegraphics[width=\linewidth]{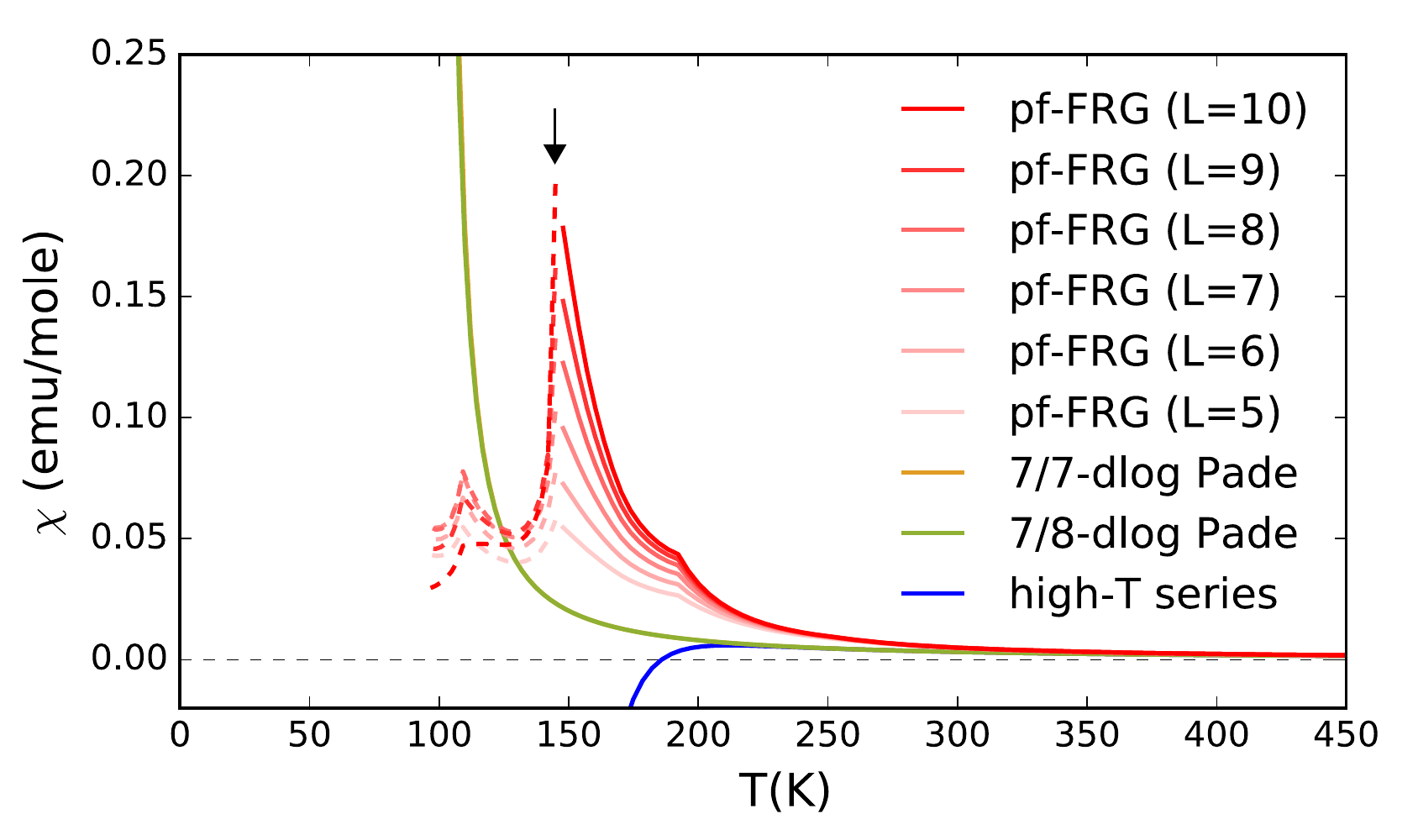}
  \caption{Susceptibility of the Heisenberg ferromagnet on the hyperkagome lattice 
  		(with a coupling strength that is assumed to be equal to the antiferromagnet). 
		The arrow indicates the breakdown point of the flow upon onset of ferromagnetism. }
  \label{fig:hyperkagomeFMSusceptibility}
\end{figure}

We round off our discussion by considering the Heisenberg {\it ferromagnet} on the hyperkagome lattice, which is of course expected to undergo a finite-temperature phase transition to a magnetically ordered state. Considering pf-FRG data for the uniform susceptibility obtained for different system sizes, we identify the location of this transition with the point at which the RG flow of the susceptibility breaks down as illustrated by the arrow in Fig.~\ref{fig:hyperkagomeFMSusceptibility}. This breakdown of the RG flow at the magnetic transition traces back 
to the fact that we enforce SU(2) spin-rotational symmetry in the flow equations at all times (in order to keep the computational effort manageable); at the magnetic transition it is, of course, precisely this SU(2) spin-rotational symmetry that is spontaneously broken and as a consequence leads to a disruption of the hitherto smooth RG flow. Although we cannot cross the phase transition into the magnetically ordered phase we can still locate the magnetic transition with high accuracy and determine the type of developing magnetic order from the static $k$-space-resolved magnetic susceptibility just above the transition (see the next Section).
Returning to the pf-FRG data of Fig.~\ref{fig:hyperkagomeFMSusceptibility}, we again find a systematic finite-size behavior. Within the resolution of our frequency/temperature discretization, the breakdown of the flow occurs at the same frequency/temperature point, but the clearly noticeable growth of the susceptibility with increasing system size indicates a slight movement of the divergence to higher temperatures for larger system sizes.
Comparing our pf-FRG results with the high-temperature series expansion, we again find the pf-FRG data to closely track the series results down to the temperature regime where the bare series is found to diverge (roughly at 200~K) and good quantitative agreement for the actual physical divergence of the susceptibility recovered from Dlog Pade approximants of the series.


\begin{figure*}[th!]
  \centering
  \includegraphics[width=\linewidth]{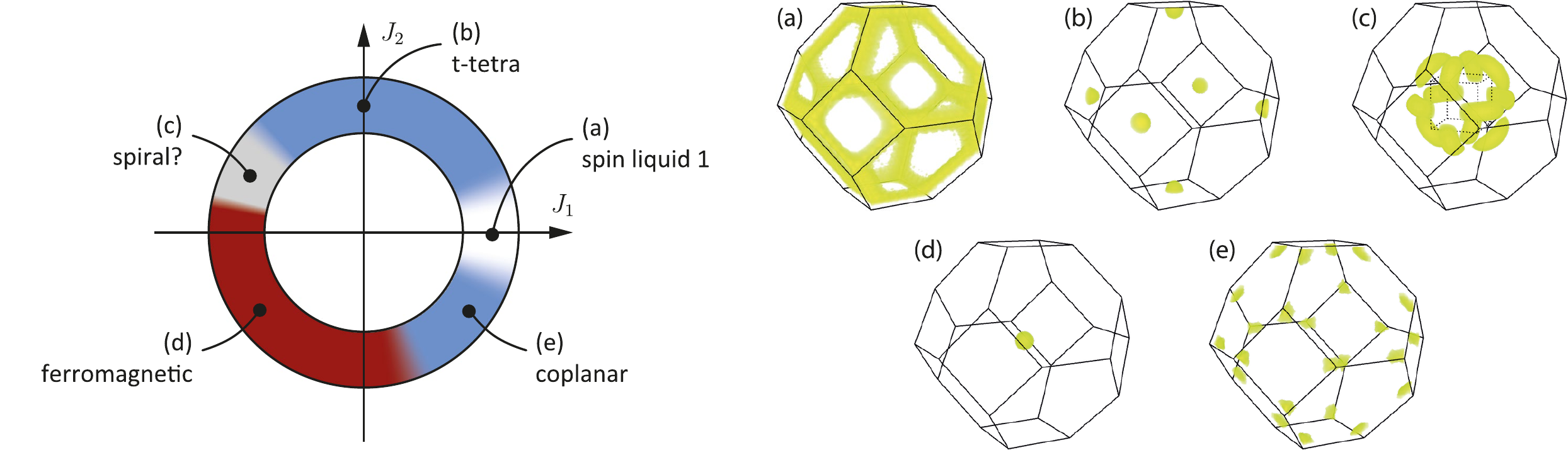}
  \caption{ Left panel: Ground-state phase diagram spanned by nearest and next-nearest neighbor couplings $J_1$ and $J_2$ ($J_3\equiv0$) on the hyperkagome lattice. Color gradients signal numerical uncertainty. Right panel: Subplots (a)-(e) display the structure factors of the different magnetically ordered phases within the extended Brillouin zone. Colored regions mark momentum positions of highest correlation. If helpful, the first Brillouin zone is indicated by dotted lines. }
  \label{fig:j1j2Hyperkagome}
\end{figure*}

\begin{figure*}[th!]
  \centering
  \vskip 3mm
  \includegraphics[width=\linewidth]{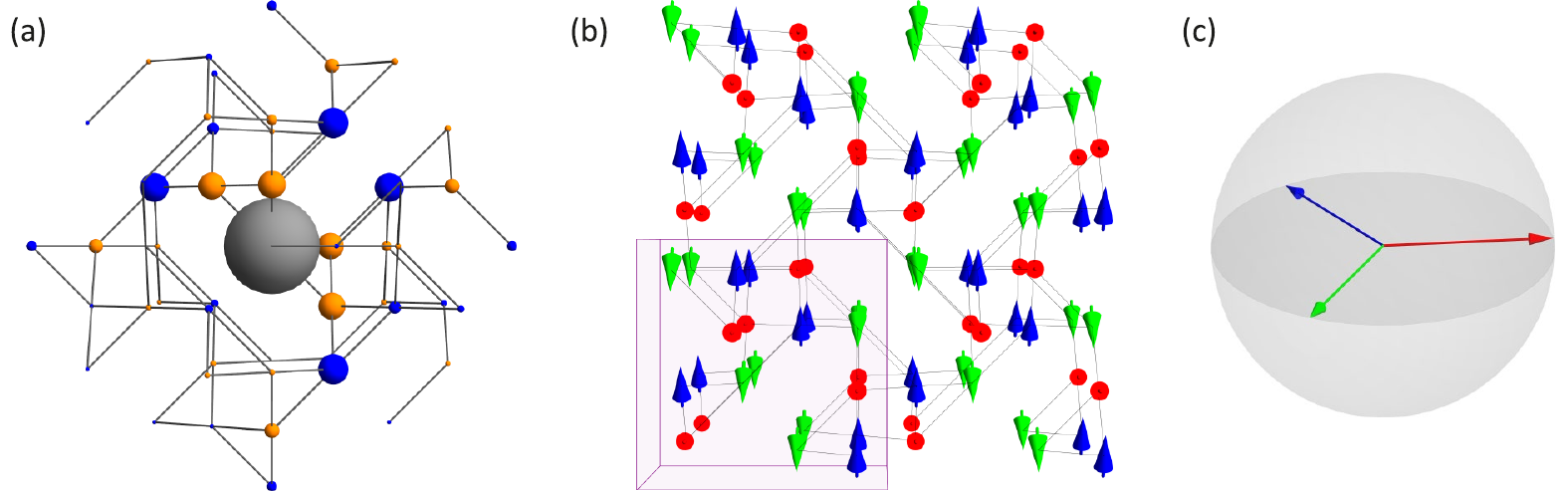}
  \caption{ The {\bf coplanar phase} of the $J_1J_2$-Heisenberg model on the hyperkagome lattice:
  		a) Real space correlations. Blue circles represent ferromagnetic correlation and orange circles AFM correlation. 
		     The correlation strength is encoded in the circles' magnitudes relative to the reference site indicated in grey.
		b) Spin configuration in real space. Indicated in purple is the magnetic unit cell which, in this phase, matches the lattice unit cell. 
			The viewpoint is along the [1 0 0]-direction. 
		c) Relevant axes of magnetic order.
		}
  \label{fig:hyperkagome-coplanar}
\end{figure*}

\section{Hyperkagome systems}
\label{sec:hyperkagome}

We now turn to a discussion of the magnetic orders and (additional) spin liquid regimes induced by next-nearest and third-nearest neighbor couplings introduced in the generalized Heisenberg model \eqref{eq:j1j2j2bHamiltonian}. Let us first consider the hyperkagome system and turn to the kagome system in the next Section.

\subsection{J$_1$-J$_2$ model}
\label{sec:hyperkagomeJ1J2}

As a first step we consider the pure Heisenberg model augmented by a next-nearest neighbor coupling along the angled bonds, which we parametrize by the coupling strength $J_2$ and therefore refer to this model also as the $J_1J_2$ model. We calculate the
phase diagram for arbitrary sign (i.e. antiferromagnetic and ferromagnetic) and relative strength of the couplings $J_1$ and $J_2$ 
by parametrizing the two couplings by an angle $\alpha\in[0,2\pi)$ as $J_1=\cos(\alpha)$ and $J_2=\sin(\alpha)$. 
For the two-dimensional kagome system it is well known that such a next-nearest neighbor coupling can stabilize a variety of magnetic orders (depending on the relative coupling strength) and we expect similar physics to be at play also for the three-dimensional hyperkagome system at hand. To identify possible magnetic orders we calculate the static $k$-space-resolved magnetic susceptibility
\begin{equation}
\chi(\mathbf{k})=\frac{1}{N}\sum\limits_{i,j}\mathrm{e}^{i\mathbf{k}(\mathbf{R}_i-\mathbf{R}_j)}\langle S^z_i S^z_j \rangle 
\end{equation}
that can again be rewritten in terms of the pseudo-fermions and therefore extracted from pf-FRG calculations in a straightforward manner. A specific magnetic ordering leads to a characteristic divergence pattern of the $k$-space-resolved magnetic susceptibility in momentum space that typically allows to reconstruct the real-space magnetic ordering. Practically, we do so by complementing the pf-FRG calculations of the $k$-space-resolved magnetic susceptibility just above the magnetic ordering transition and the associated breakdown of the RG flow (see previous Section) with classical Monte Carlo simulations for coupling parameters deep in the ordered phase and verify that the real-space magnetic ordering identified by the Monte Carlo approach gives a matching momentum-space magnetic susceptibility.

\begin{figure*}[th!]
  \centering
  \includegraphics[width=1\linewidth]{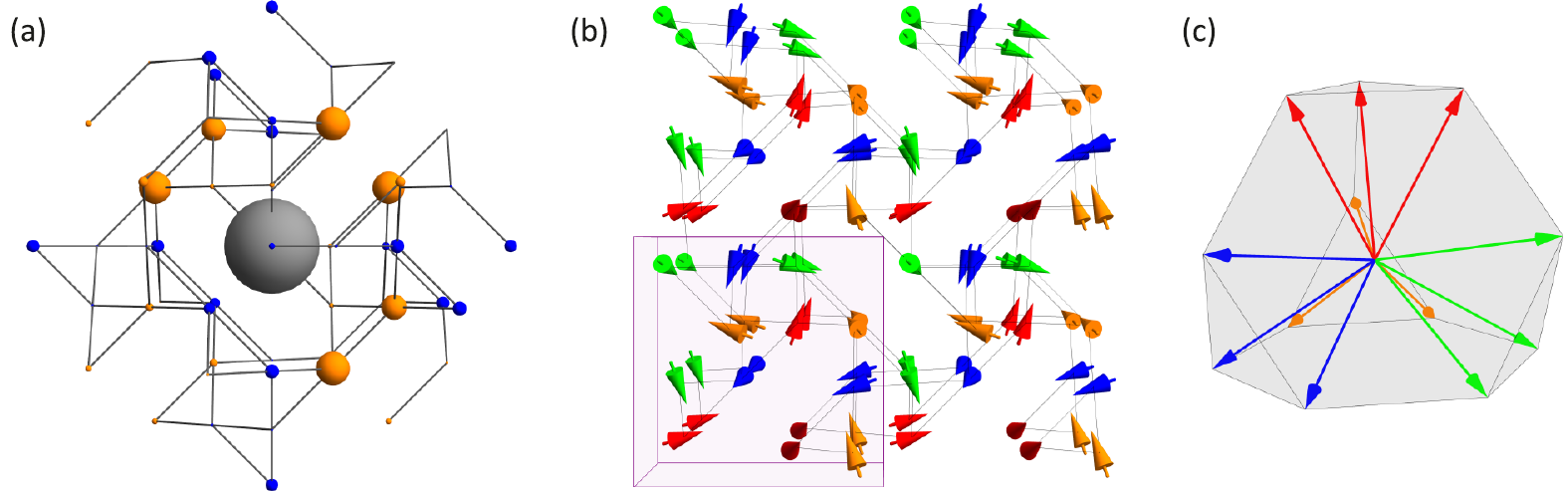}
  \caption{ The {\bf t-tetra phase} of the $J_1J_2$-Heisenberg model on the hyperkagome lattice:
  		a) Real space correlations. Blue circles represent ferromagnetic correlation and orange circles AFM correlation. 
		     The correlation strength is encoded in the circles' magnitudes relative to the reference site indicated in grey.
		b) Spin configuration in real space. 
		c) Relevant axes of magnetic order.}
  \label{fig:hyperkagome-t-tetra}
\end{figure*}

\begin{figure}[th!]
  \centering
  \includegraphics[width=0.65\linewidth]{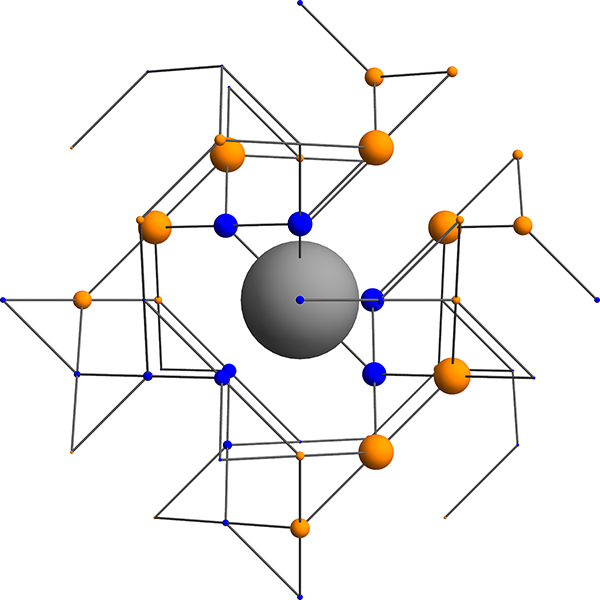}
  \caption{ Real space correlations for the weakly ordered {\bf spiral phase} of the $J_1J_2$-Heisenberg model on the hyperkagome lattice.
  		Blue circles represent ferromagnetic correlation and orange circles AFM correlation. 
		   The correlation strength is encoded in the circles' magnitudes relative to the reference site indicated in grey.
		}
  \label{fig:hyperkagome-spiral}
\end{figure}

This procedure allows us to map out the circular phase diagram of competing magnetic orders in the  $J_1J_2$ hyperkagome model parametrized by the angle $\alpha$ as shown in the left panel of Fig.~\ref{fig:j1j2Hyperkagome}. We find a total of five different phases, four of which are magnetically ordered. The simplest magnetic order is ferromagnetism which spans the entire lower left quadrant where both coupling constants are ferromagnetic. It is easily identified by its characteristic structure factor whose central peak signals the alignment of all spins in the system, see Fig.~\ref{fig:j1j2Hyperkagome}(d). The ferromagnetic phase persists if either $J_1$ or $J_2$ turns slightly antiferromagnetic but as soon as the AFM interactions become sizable a new magnetic ground state with a more subtle magnetic texture is favored. In the case of antiferromagnetic $J_1$ the resulting state is a coplanar $120^\circ$ order with a magnetic unit cell that is of the same size as the lattice unit cell as illustrated in Fig.~\ref{fig:hyperkagome-coplanar}. The associated structure factor shown in Fig.~\ref{fig:j1j2Hyperkagome}(e) shows clear peaks in the corners of the extended Brillouin zone. If we assess the coplanar order not by its structure factor but by its real space resolved spin correlations, shown in the left panel of Fig.~\ref{fig:hyperkagome-coplanar}, one can immediately understand why this particular order is favored -- it indeed shows strong antiferromagnetic correlations on the nearest neighbor level and strong ferromagnetic correlations of next-nearest neighbors. A different picture arises when we consider ferromagnetic $J_1$ with moderate antiferromagnetic next-nearest neighbor interaction. In that setting our pf-FRG 
simulations point to weak, likely non-coplanar magnetic ordering, possibly the formation of spin spirals.
The structure factor, shown in Fig.~\ref{fig:j1j2Hyperkagome}(c), exhibits smeared-out features around the first Brillouin zone indicative of
multiple degenerate magnetic orderings with a relatively large magnetic unit cell. The real-space correlations confirm ferromagnetic alignment of nearest neighbors although the correlation is comparably weak, see Fig.~\ref{fig:hyperkagome-spiral}. 
The combination of  ferromagnetic correlations amongst nearest neighbors and strong antiferromagnetic correlations between next-nearest neighbors (as favored by $J_2$ in this part of the phase diagram) likely gives rise to a non-coplanar spin order. 
Further characterizing the precise magnetic order in this phase has turned out to be unexpectedly tedious, as our supplementary classical Monte Carlo simulations do not find a magnetic ordering pattern that reproduces the features of the structure factor found in the pf-FRG calculations -- pointing to the possibility that quantum fluctuations stabilize the weak magnetic order seen in the pf-FRG simulations.
In the upper half of the phase diagram a fourth magnetically ordered phase forms in an extended range of parameters that exhibits antiferromagnetic correlations not just for next-nearest neighbors (as favored by the positive $J_2$ in these quadrants) but also for nearest neighbors (independent of the sign of the nearest neighbor coupling). In order to form antiferromagnetic correlations of this sort the twelve spins in the (magnetic) unit cell align such that they span a truncated tetrahedron as illustrated in the right panel of Fig.~\ref{fig:hyperkagome-t-tetra}. We therefore refer to this phase also as the ``t-tetra phase". The real-space correlations of this phase are illustrated in the left panel of Fig.~\ref{fig:hyperkagome-t-tetra}. The corresponding magnetic structure factor has distinct peaks on the square surface segments of the extended Brillouin zone as shown in Fig.~\ref{fig:j1j2Hyperkagome}(b). 

Beyond these four magnetically ordered phases we find indications of an extended spin liquid phase in a small wedge of parameter space around purely antiferromagnetic nearest-neighbor exchange $J_1=1$, i.e. $\alpha = 0$. The finite-temperature signatures of the magnetic susceptibility for this phase are in striking agreement with the available experimental data for the hyperkagome spin liquid candidate material Na$_4$Ir$_3$O$_8$ as discussed in the Introduction. Adding a small next-nearest neighbor coupling $J_2$ (of either sign) slightly shifts the magnetic susceptibility, but cannot explain the small upturn around 20~K observed in the experimental data (see Fig.~\ref{fig:hyperkagomeAFMSusceptibilityExperiment}). In the following, we concentrate on the low-temperature signatures of this phase.
 
\begin{figure}[t]
  \centering
  \includegraphics[width=0.65\linewidth]{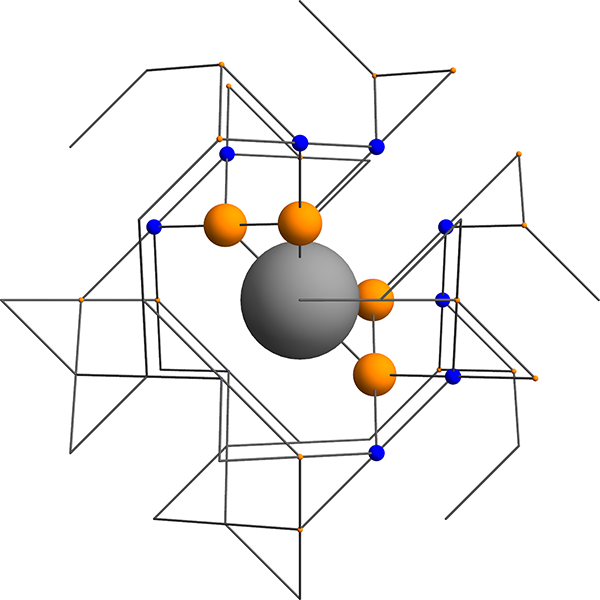} \\
  \includegraphics[width=\linewidth]{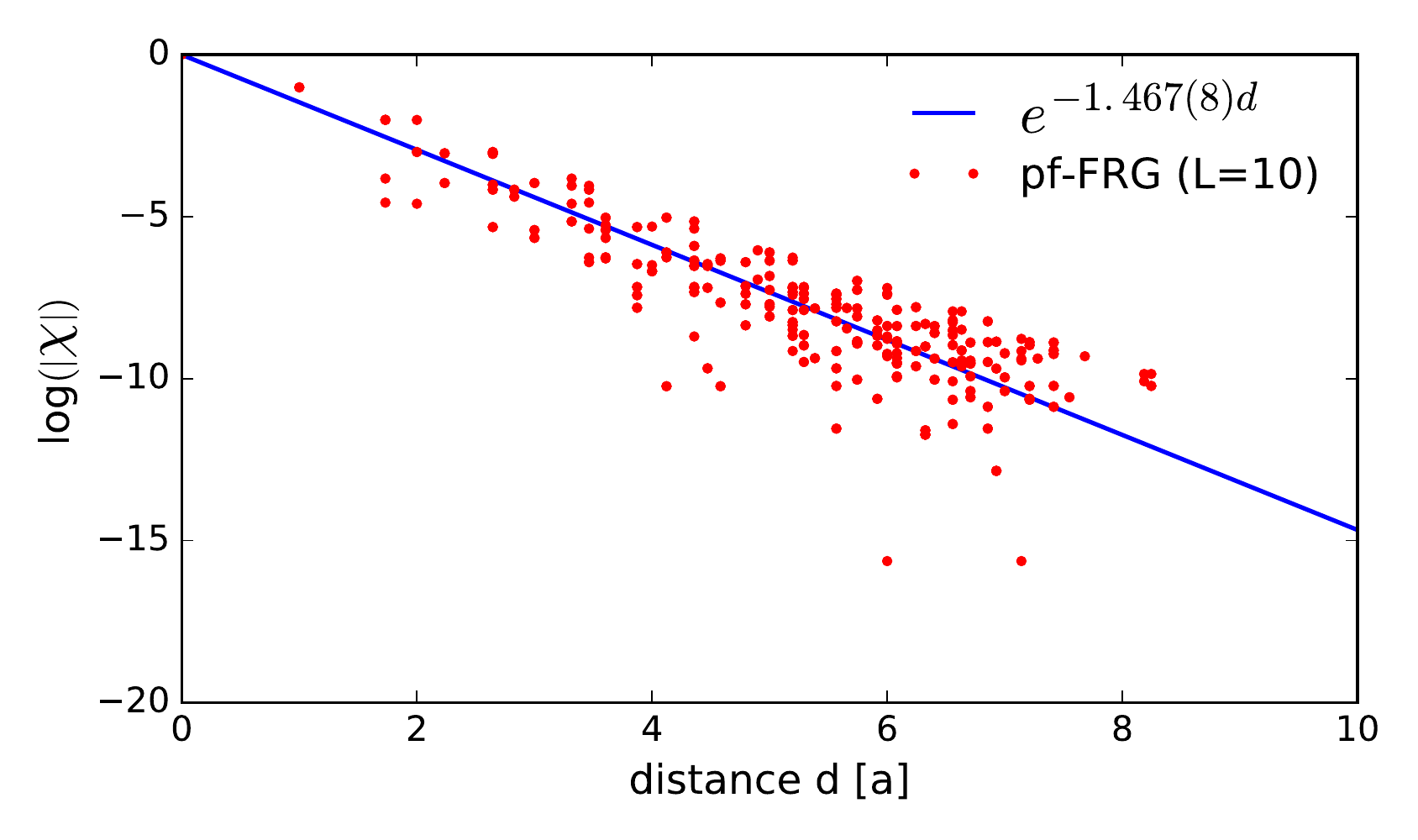}
  \caption{Real space correlations for the {\bf spin liquid 1 phase} of the $J_1J_2$-Heisenberg model on the hyperkagome lattice. 
		Top panel: Blue circles represent ferromagnetic correlation and orange circles AFM correlation. 
		The correlation strength is encoded in the circles' magnitudes relative to the reference site indicated in grey. 
		Bottom panel: Decay of real space correlations as a function of distance in units of the lattice constant $a$. 
		The pf-FRG result (red) is fitted by an exponential decay. }
  \label{fig:hyperkagome-SL1}
\end{figure}

Rigorously establishing the nature of such a putative spin liquid phase is a difficult task within the pf-FRG framework. In fact, the approach relies on identifying the {\it absence} of any imaginable magnetic order, i.e. the absence of a breakdown of the RG flow due to breaking spin rotational symmetry (see the discussion at the end of Section \ref{sec:pfFRG} above). While such an indirect approach to the diagnosis of potential spin liquid phases has long been pursued, it has become somewhat of a substandard approach in recent years given that a positive identification of spin liquids via their macroscopic entanglement structure \cite{KitaevPreskill2006,LevinWen2006}, topological invariants \cite{Vishwanath2012,Cincio2013}, or inherent gauge structure has been achieved in other numerical approaches \cite{Jiang2012,Bauer2014,He2014,Gong2014,Nasu2014}. In addition, the general setup of the pseudofermion approach seems to have an inherent preference towards spin liquid phases with a U(1) gauge structure that preserve the spin rotational symmetry enforced in the RG equations. Other gauge structures which are encountered in Z$_2$ spin liquids or chiral spin liquids are likely to be missed in this setup and would presumably require a different type of spin decomposition \eqref{eq:pseudofermions} such as the well-known Majorana fermionization in Kitaev spin liquids \cite{Kitaev2006}.

Having said these words of caution we return to the putative spin liquid phase in the hyperkagome model at hand. One way to further verify and characterize the spin liquid nature is to consider its real-space correlations, which as shown in Fig.~\ref{fig:hyperkagome-SL1} reveal extremely fast decaying correlations -- a scenario more in line with a gapped Z$_2$ spin liquid than a U(1) spin liquid for which we expect algebraically decaying correlation functions. Further testament to the spin liquid nature is provided by the spin structure factor plotted in Fig.~\ref{fig:j1j2Hyperkagome}(a) exhibiting smeared out features around the edges of the Brillouin zone -- a feature that is in striking resemblence of the kagome spin liquid, which we will discuss in Section \ref{sec:kagome} below.


\subsection{J$_1$-J$_{2=3}$ model}
\label{sec:hyperkagomeJ1J23}

\begin{figure}[b]
  \centering
  \includegraphics[width=0.7\linewidth]{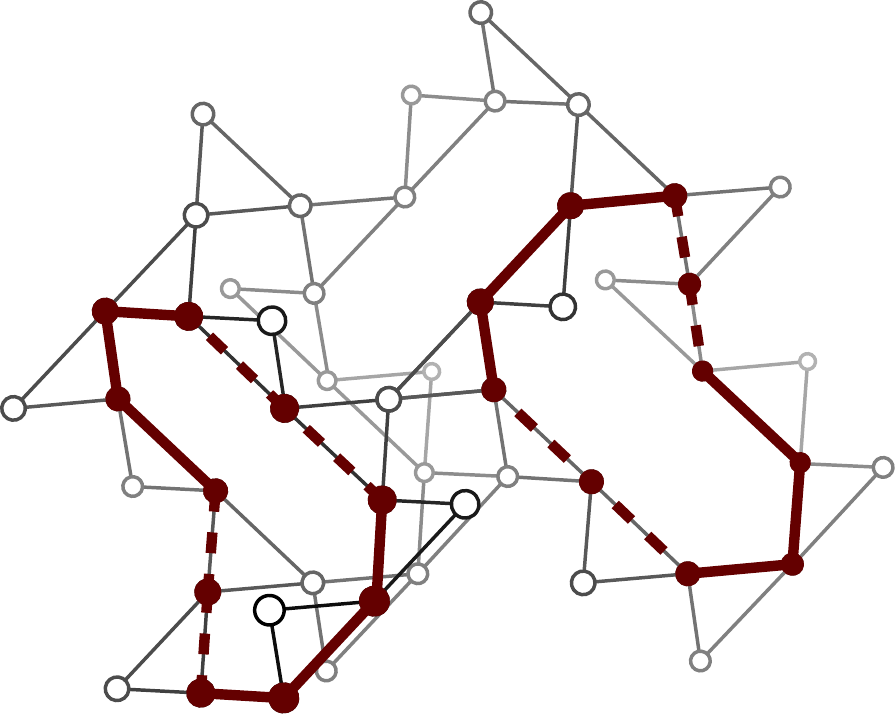}
  \caption{ Elementary loops on the hyperkagome lattice (solid lines). Every loop involves two pairs of third-nearest neighbors (dashed line) that interact via the exchange constant $J_3$. }
  \label{fig:hyperkagomeLoops}
\end{figure}

\begin{figure*}[th]
  \centering
  \includegraphics[width=\linewidth]{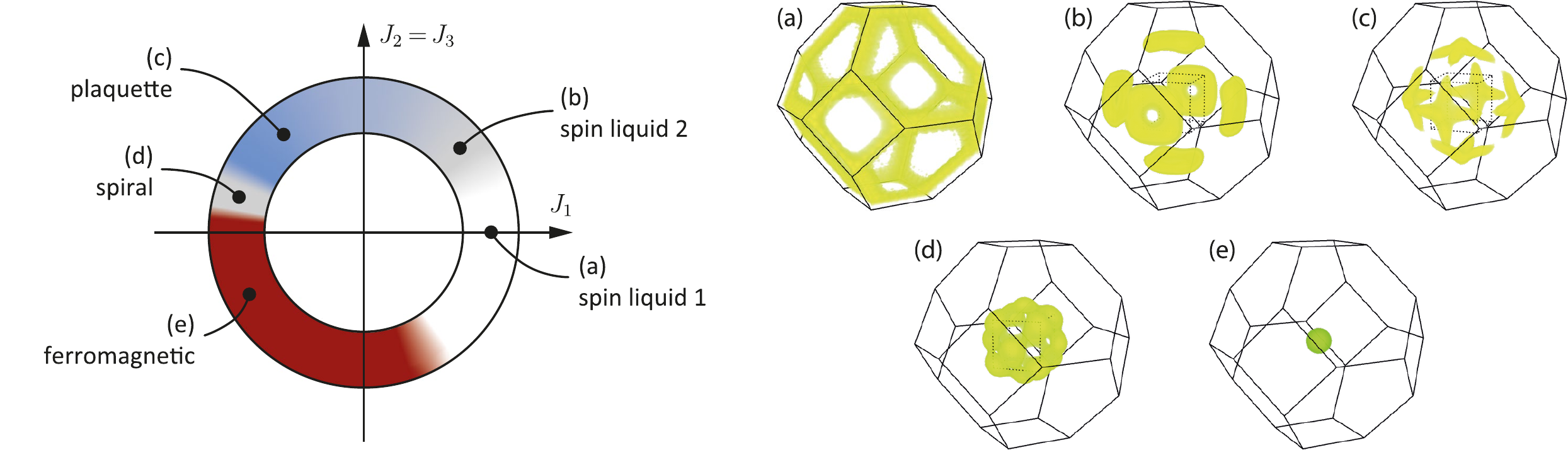}
  \caption{ Left panel: Phase diagram of the $J_1J_{2=3}$-Heisenberg model on the hyperkagome lattice for equal strength second- and third-nearest neighbor couplings $J_2=J_3$. Right panels: structure factors of the various phases plotted within the extended Brillouin zone.}
  \label{fig:j1j2bHyperkagome}
\end{figure*}

Taking a step back one might ask why there is no second spin liquid regime in the $J_1J_2$ model on the hyperkagome lattice at and around purely antiferromagnetic next-nearest neighbor coupling $J_2=1$, i.e. at and around $\alpha = \pi / 2$ in the circle phase diagram. This question is informed by the fact that the elementary plaquette of the hyperkagome lattice (beyond the triangles) is a decagon, which is constituted by a 10-bond loop as illustrated in Fig.~\ref{fig:hyperkagomeLoops}. Naturally, an antiferromagnetic coupling of all next-nearest neighbors along such a loop would seem to induce geometric frustration along the two effective pentagons formed by the next-nearest neighbor spins. Taking a closer look, however, reals that any such pentagon is constituted by four angled bonds and one collinear third-nearest neighbor bond. Since our $J_1J_2$ model included only the angled next-nearest neighbor bonds, this elementary loop structure was never closed and no geometric frustration encountered.
It is thus a natural question to ask what physics is induced when including the collinear third-nearest neighbor bonds. To do so, we consider a $J_1J_{2=3}$ model, in which the next-nearest and third-nearest neighbor couplings enter at the same strength, i.e. $J_2 = J_3$. This choice of equal couplings is partially motivated by the observation that the microscopic origin of these two couplings along the two bonds is rather similar with a dominant Ir-O-Ir-O-Ir exchange path along the two bonds in Na$_4$Ir$_3$O$_8$. In practice, this means that  each site that previously had 6 second-nearest neighbors is now coupled to an additional 2 third-nearest neighbors with coupling $J_3$. As expected by the above arguments, this seemingly small change destabilizes the magnetic ordering tendencies in the original $J_1J_2$ model and results in a drastically different phase diagram as illustrated in Fig.~\ref{fig:j1j2bHyperkagome}.

Comparing the phase diagram of the  $J_1J_{2=3}$ hyperkagome model (Fig.~\ref{fig:j1j2bHyperkagome}) to the  $J_1J_2$ hyperkagome model (Fig.~\ref{fig:j1j2Hyperkagome}), probably the most obvious change is the disappearance of the magnetically ordered coplanar and t-tetra phases for the  $J_1J_{2=3}$ hyperkagome model. Let us first consider the disappearance of the coplanar phase which can easily be rationalized by considering its  real-space correlations illustrated in the left panel of Fig.~\ref{fig:hyperkagome-coplanar}. Close inspection reveals that the 2 third-nearest neighbors align antiferromagnetically in this phase, which conflicts with a ferromagnetic $J_3$ coupling in this quadrant of the phase diagram of the  $J_1J_{2=3}$ hyperkagome model. As a consequence, the coplanar order is destabilized and vanishes entirely from the phase diagram. In its place the spin liquid of the pure hyperkagome antiferromagnetic ($\alpha=0$) takes over and now extends over a wide range of parameter space.

\begin{figure}[h!]
  \centering
  \includegraphics[width=0.65\linewidth]{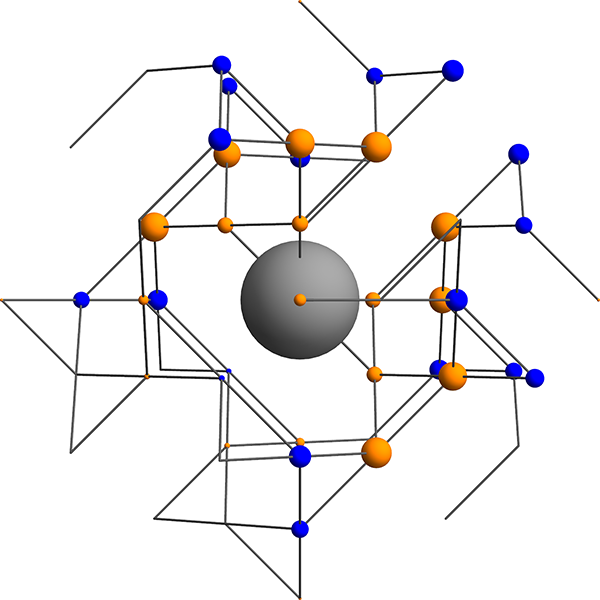}
  \includegraphics[width=\linewidth]{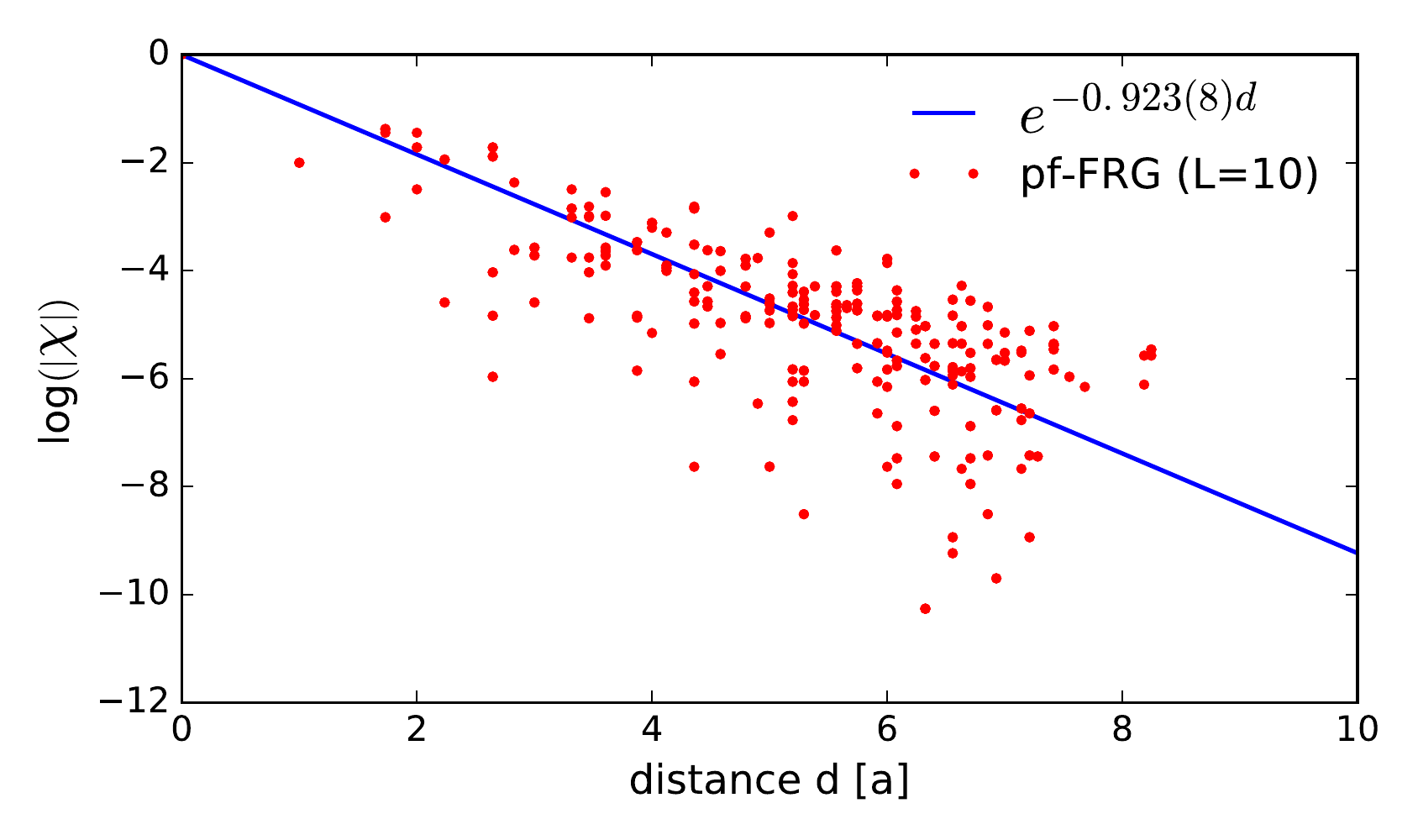}
  \caption{Real space correlations for the {\bf spin liquid 2 phase} of the $J_1J_{2=3}$-Heisenberg model on the hyperkagome lattice.
		Top: Blue circles represent ferromagnetic correlation and orange circles AFM correlation. 
		The correlation strength is encoded in the circles' magnitudes relative to the reference site indicated in grey. 
		Bottom panel: Decay of real space correlations as a function of distance in units of the lattice constant $a$. 
		The pf-FRG result (red) is fitted by an exponential decay. }
  \label{fig:hyperkagome-SL2}
\end{figure}

As argued above the geometric frustration induced by the combination of second- and third-nearest neighbor couplings $J_2$ and $J_3$ along the elementary 10-bond loop leads to a destabilization of the t-tetra magnetically ordered phase in the upper half of the phase diagram of the  $J_1J_2$ model. Similar to the coplanar ordering this vanishing of t-tetra order can also be rationalized by the observation that an antiferromagnetic third-neighbor coupling $J_3$ is conflicting with the ferromagnetic correlations of third-nearest neighbors in the  t-tetra ordered phase, see its real-space correlations in the left panel of Fig.~\ref{fig:hyperkagome-t-tetra}.
Penalizing these correlations with the third-nearest neighbor couplings instead gives rise to a second type of spin liquid phase, which is clearly distinct from the spin liquid phase of the hyperkagome antiferromagnet ($\alpha=0$).
Although our pf-FRG calculations do not allow us to classify spin liquids in terms of their symmetries we do observe a sudden qualitative change in the structure factor at $\alpha\approx\pi/6$, i.e. for $J_2 = J_3 \approx 0.58~J_1$. The spin liquid phase around the hyperkagome antiferromagnet ($\alpha=0$) exhibits a structure factor with broad features on the edges of the extended Brillouin zone in the entire parameter range indicated in the phase diagram of Fig.~\ref{fig:j1j2bHyperkagome}. In contrast, the second spin liquid shows broadly smeared out features (not providing any evidence for the formation of peaks indicating magnetic order) no longer situated on the edges of the extended Brillouin zone but considerably closer to the first Brillouin zone, see the illustration in Fig.~\ref{fig:j1j2bHyperkagome}(b). This apparent collapse of the spin-liquid features to smaller momentum scales indicates an expansion of the relevant real space scales, which we expect once frustration on (large) 10-bond loops outweighs the frustration on (small) triangular plaquettes. The real-space spin correlations of this second spin liquid phase are depicted in Fig.~\ref{fig:hyperkagome-SL2}, which -- in comparison with the first spin liquid around the pure hyperkagome antiferromagnet ($\alpha=0$), see Fig.~\ref{fig:hyperkagome-SL1} --  reveal a somewhat slower decay of magnetic correlations, which however still seems to be exponential (see the fit in the lower panel of Fig.~\ref{fig:hyperkagome-SL2}) indicative of a gapped spin liquid.
 
\begin{figure*}
  \centering
  \includegraphics[width=\linewidth]{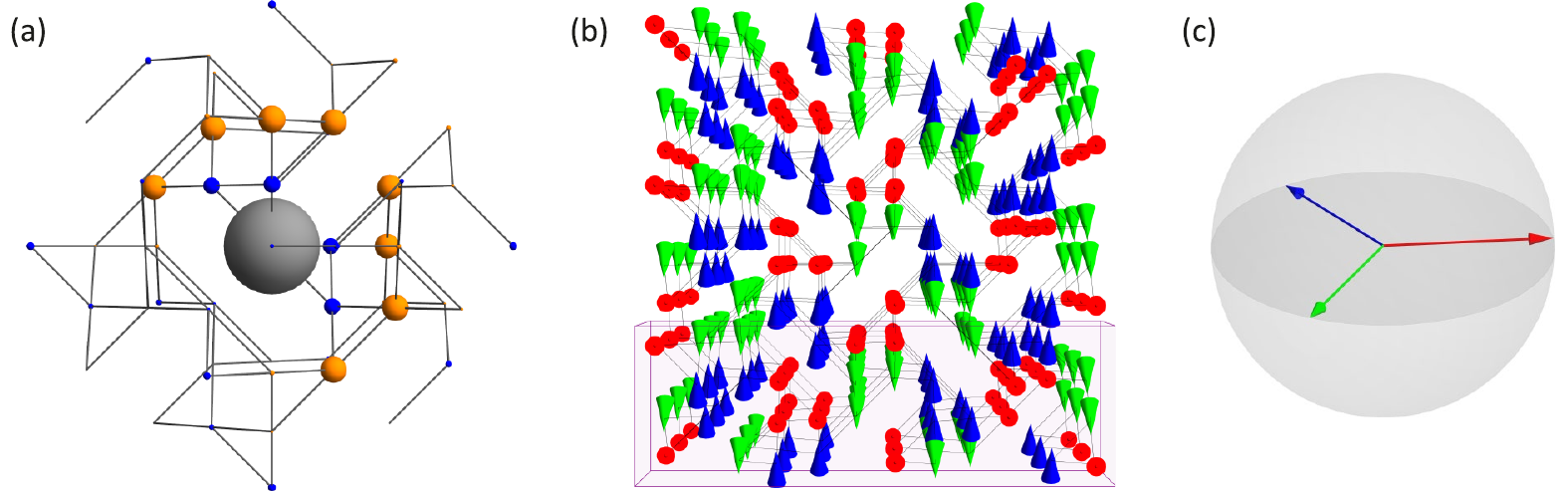}
    \caption{ The {\bf plaquette phase} of the $J_1J_{2=3}$-Heisenberg model on the hyperkagome lattice:
  		a) Real space correlations. Blue circles represent ferromagnetic correlation and orange circles AFM correlation. 
		     The correlation strength is encoded in the circles' magnitudes relative to the reference site indicated in grey.
	        b) Spin configuration in real space. 
		c) Relevant axes of magnetic order.
		}
  \label{fig:hyperkagome-plaquette}
\end{figure*}

Relieving the system from the short-range geometric frustration by turning the nearest-neighbor coupling $J_1$ ferromagnetic is found to induce a novel form of magnetic ordering, which we dub plaquette order. It is a coplanar $120^\circ$-degree spin ordering that is coarse-grained such that spins on stacked triangular plaquettes align ferromagnetically as illustrated in the middle panel of Fig.~\ref{fig:hyperkagome-plaquette}. An analysis of its real space correlations (left panel in Fig.~\ref{fig:hyperkagome-plaquette}) reveals that this ordering is fully compatible with ferromagnetic nearest neighbor and antiferromagnetic next-nearest neighbor interactions.
The associated structure factor, depicted in Fig.~\ref{fig:j1j2bHyperkagome}(c), stands out because unlike for the other magnetic phases it does not reveal singular peaks but slightly broader features. This peculiar form of the structure factor can be traced back to the unconventional form of its  $3\times1\times1$ magnetic unit cell which blurs out one spatial dimension. As a consequence, the structure factor of a single such spin configuration is slightly asymmetric. Upon symmetrization one therefore obtains several close-lying maxima that form a cross-shape with a slightly elevated center. This symmetrized structure factor is indeed what we observe in our pf-FRG calculations. Because of these almost surface-like peaks the precise phase boundary to the spin liquid 2 is difficult to determine (and indicated by the shaded transition in the phase diagram of Fig.~\ref{fig:j1j2bHyperkagome}). In the pf-FRG calculation we merely observe a smooth evolution of the structure factor. 

Finally, we turn to  the slim region between the plaquette phase and the ferromagnetic phase which is denoted as spiral phase in the phase diagram of Fig.~\ref{fig:j1j2bHyperkagome}. The spiral phase arises from what can be considered a smooth transition between the plaquette and ferromagnetic phases. Starting at $\alpha \approx 0.89\pi$ the features of the structure factor continuously shrink towards the center of the Brillouin zone until they become just a single peak which marks the onset of ferromagnetism. This scenario is very much akin to the physics of the two-dimensional kagome lattice with analogous second- and third-nearest neighbor interactions as we will discuss below.

We close this Section by noting that it would be interesting to make a connection between the spin liquid phases observed for the hyperkagome systems and the physics of similar $J_1J_2J_3$ models on the {\it pyrochlore} lattice, which have recently been studied in Refs.~\onlinecite{Ishizuka2013,Udagawa2016,Rau2016}.


\begin{figure*}[th!]
  \centering
  \includegraphics[width=\linewidth]{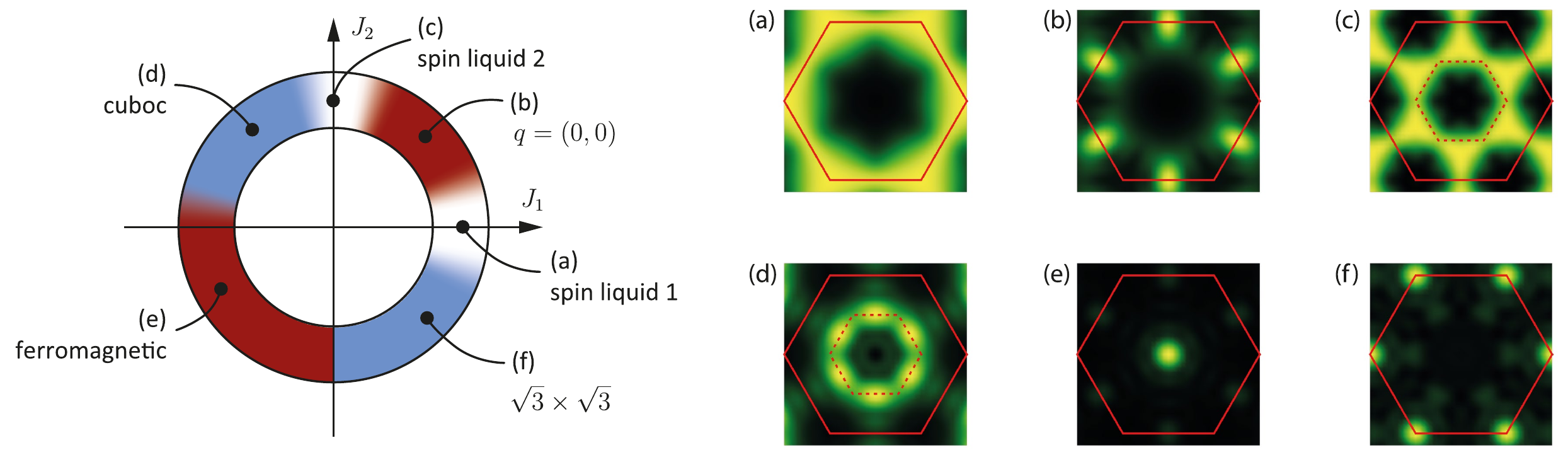}
  \caption{ Left panel: Phase diagram of the  $J_1J_2$ -Heisenberg model on the kagome lattice. 
  		Right panels: structure factors of the magnetically ordered phases (b) and (d)-(f) as well as the two spin liquid phases (a) and (c). The extended Brillouin zone is indicated by a solid red line which, if useful, is complemented by the first Brillouin zone indicated by a dotted line. }
  \label{fig:j1j2Kagome}
\end{figure*}

\begin{figure*}[th!]
  \centering
  \includegraphics[width=\linewidth]{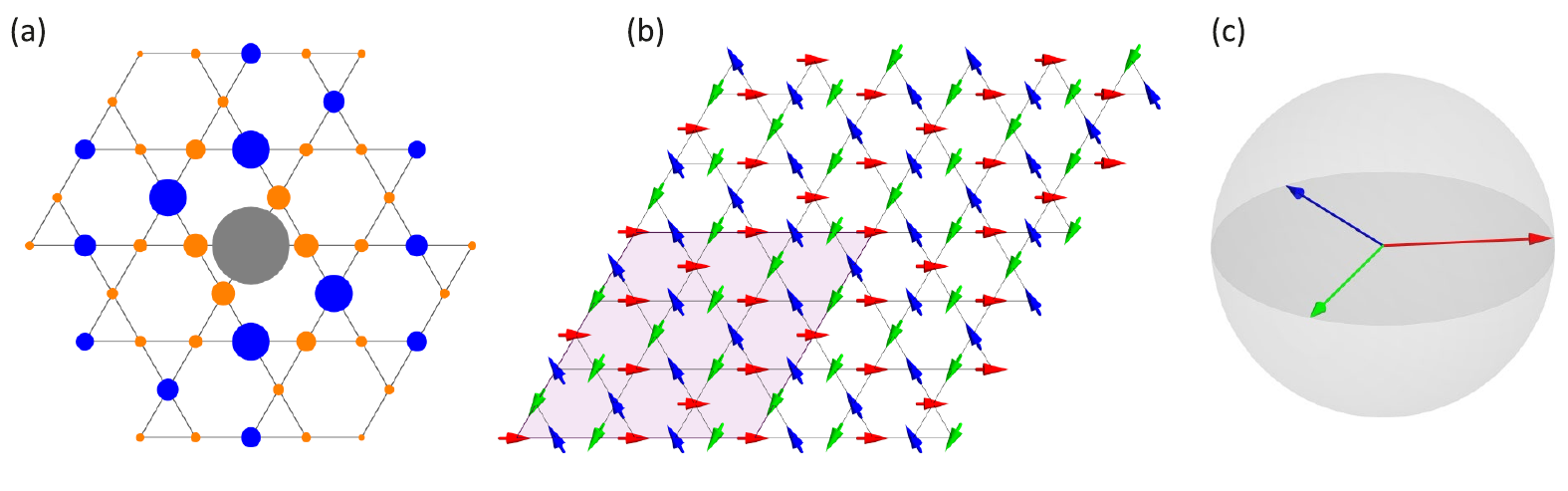}
  \caption{ The {\bf $\mathbf{\sqrt{3}\times\sqrt{3}}$  phase} of the $J_1J_2$-Heisenberg model on the kagome lattice:
  		a) Real space correlations. Blue circles represent ferromagnetic correlation and orange circles AFM correlation. 
		     The correlation strength is encoded in the circles' magnitudes relative to the reference site indicated in grey.
		b) Spin configuration in real space.
		c) Relevant axes of magnetic order.
		}
  \label{fig:kagome-sqrt3}
\end{figure*}

\section{Kagome systems}
\label{sec:kagome}

To put our results for the generalized   $J_1J_2$ and $J_1J_{2=3}$ Heisenberg models on the hyperkagome lattice into context and to  further exemplify the versatility of the pf-FRG approach, we consider analogous models for the two-dimensional kagome system. As mentioned already in the introduction the two-dimensional kagome model has received a great amount of theoretical attention as one of the most fundamental frustrated quantum magnets \cite{Diep2004}. Closest to our approach here are extensive numerical studies \cite{Lhuillier2012,Gong2015} of the $J_1J_2J_d$ kagome model with next-nearest neighbor couplings $J_2$ along angled next-nearest neighbor bonds (see Fig.~\ref{fig:lattices}) and a third-nearest neighbor coupling $J_d$, diagonally across the plaquettes of the elementary hexagons. While these previous studies cover the $J_1J_2$ model of interest here, the role of third-nearest neighbor couplings along collinear bonds -- which we parametrize by $J_3$ and should be distinguished from the diagonal cross-plaquette coupling $J_d$ \cite{Gong2015,Bieri2015} -- has, to the best of our knowledge, not been elucidated in any detail. A classical variant of this model has been studied in Ref.~\onlinecite{Messio2011}.

In the following, we will quickly go through the $J_1J_2$  model and demonstrate that the pf-FRG approach faithfully reproduces its known phase diagram. We then turn to the role of the third-nearest neighbor couplings J$_3$, which are found to destabilize all the non-trivial magnetic orders of the $J_1J_2$  model -- very much akin to the situation of the analogous hyperkagome model discussed in the previous Section.
We close with a discussion of the general $J_1J_2J_3$ model with arbitrary relative strength of the second-nearest and third-nearest neighbor couplings.


\subsection{J$_1$-J$_2$ model}
\label{sec:kagomeJ1J2}

For the sake of completeness, we start our discussion of kagome systems by applying the pf-FRG approach to the $J_1J_2$-Heisenberg model and compare the obtained phase diagram to the one of the corresponding $J_1J_2$ hyperkagome model
discussed in Section \ref{sec:hyperkagomeJ1J2}. In general, we find that our pf-FRG calculations are faithfully reproducing the known phase diagram of this model \cite{Lhuillier2012,Gong2015} and are also perfectly in line with a previous pf-FRG study \cite{Suttner2014} of this model by other authors \cite{FootnoteFRGPhaseDiagrams}.

In a nutshell, the obtained phase diagram illustrated in Fig.~\ref{fig:j1j2Kagome} includes four magnetically ordered phases (with ferromagnetic, $\sqrt{3}\times\sqrt{3}$, ${\bf q}=(0,0)$, and cuboc order) and two spin liquid regimes. Some of these phases can be related directly to an equivalent phase on the three-dimensional hyperkagome model but there are also some minor differences. Beyond the obvious relation of the two ferromagnetic phases, there is a considerable similarity between the kagome $\sqrt{3}\times\sqrt{3}$-phase, summarized in Fig.~\ref{fig:kagome-sqrt3}, and the coplanar phase of the hyperkagome model. Both phases are based on coplanar $120^\circ$-degree order with spins ordering ferromagnetically within each of the three sublattices. This similarity is also reflected in the respective structure factors, which for both phases exhibit characteristic peaks in the corners of the extended Brillouin zone. 

\begin{figure*}[th!]
  \centering
  \includegraphics[width=\linewidth]{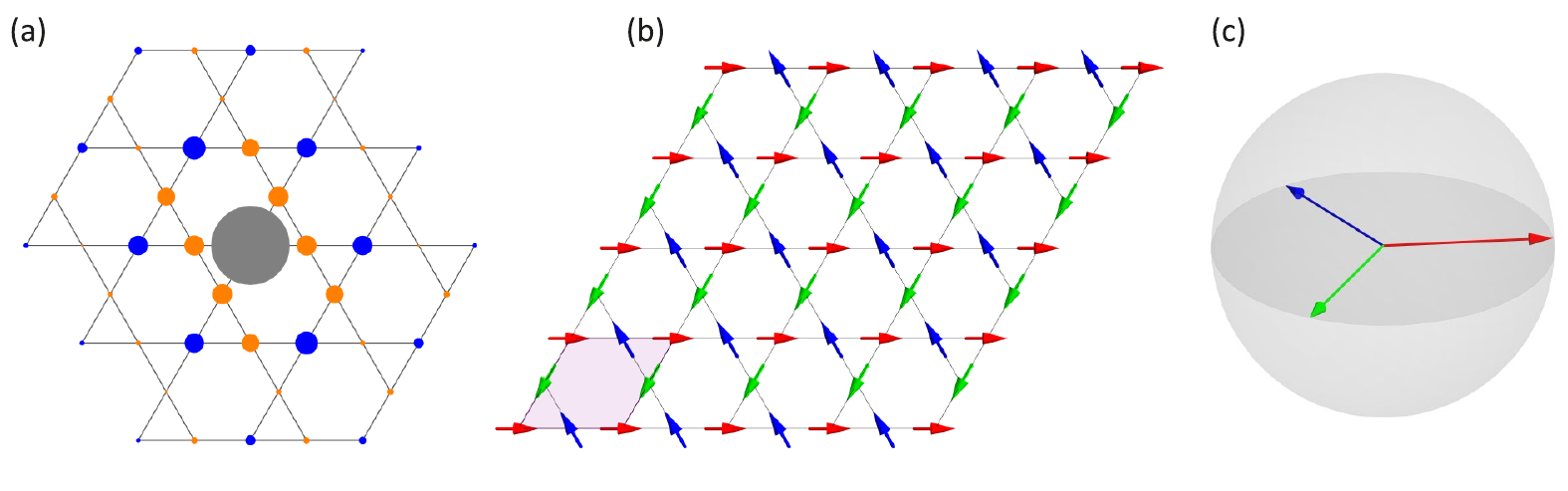}
  \caption{ The {\bf $\mathbf{q}=(0,0)$  phase} of the $J_1J_2$-Heisenberg model on the kagome lattice:
  		a) Real space correlations. Blue circles represent ferromagnetic correlation and orange circles AFM correlation. 
		     The correlation strength is encoded in the circles' magnitudes relative to the reference site indicated in grey.
		b) Spin configuration in real space. 
		c) Relevant axes of magnetic order.
		}
  \label{fig:kagome-q0}
\end{figure*}

\begin{figure*}[th!]
  \centering
  \includegraphics[width=\linewidth]{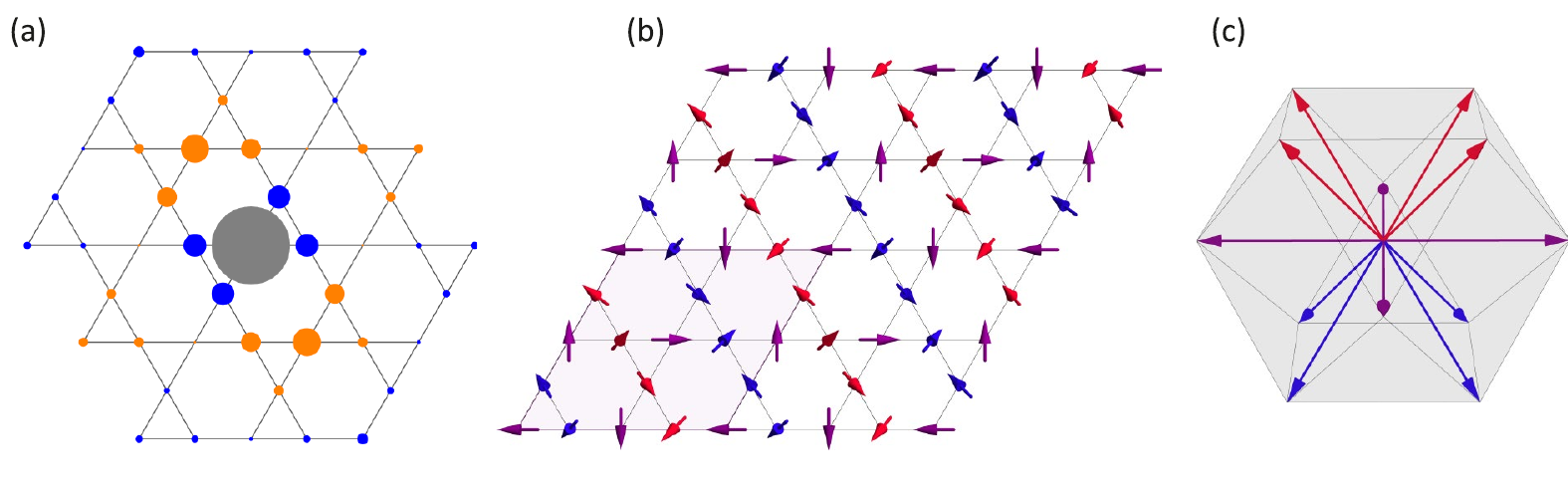}
  \caption{ {\bf Cuboc phase} of the $J_1J_2$-Heisenberg model on the kagome lattice:
  		a) Real space correlations. Blue circles represent ferromagnetic correlation and orange circles AFM correlation. 
		     The correlation strength is encoded in the circles' magnitudes relative to the reference site indicated in grey.
		b) Spin configuration in real space. 
		c) Relevant axes of magnetic order.}
  \label{fig:kagome-cuboc}
\end{figure*}

A certain similarity of their structure factors also connects the ${\bf q}=(0,0)$ ordered phase of the kagome model and the t-tetra phase of the hyperkagome system, yet there are also some differences. For the kagome phase the structure factor is dominated by peaks on the surface of the extended Brillouin zone. Similarly, for the  hyperkagome phase the strongest peaks are found only on the square faces of the Brillouin zone, with additional peaks on the hexagonal faces that are somewhat weaker. This seemingly small difference becomes more striking when comparing the actual spin configurations -- the ${\bf q}=(0,0)$ kagome order is coplanar, while the t-tetra hyperkagome order is non-coplanar.

The cuboc phase is the only non-coplanar phase for the $J_1J_2$ kagome model. Its name stems from the twelve different spin orientations that span a cuboctahedron within the magnetic unit cell, which at a size of $2\times2$ lattice unit cells is relatively large. The structure factor of the kagome cuboc order features peaks on the edges of the first Brillouin zone, see Fig.~\ref{fig:j1j2Kagome}(d) which is  similar to what we have seen for the spiral phase on the hyperkagome lattice occupying the same parameter space in the 
$J_1J_2$  hyperkagome model, see Fig.~\ref{fig:j1j2Hyperkagome}(c). 

\begin{figure}
  \centering
  \includegraphics[width=0.65\linewidth]{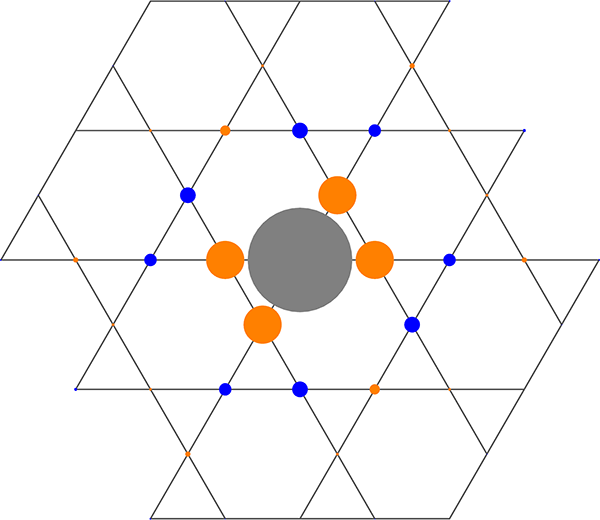} \\
  \includegraphics[width=\linewidth]{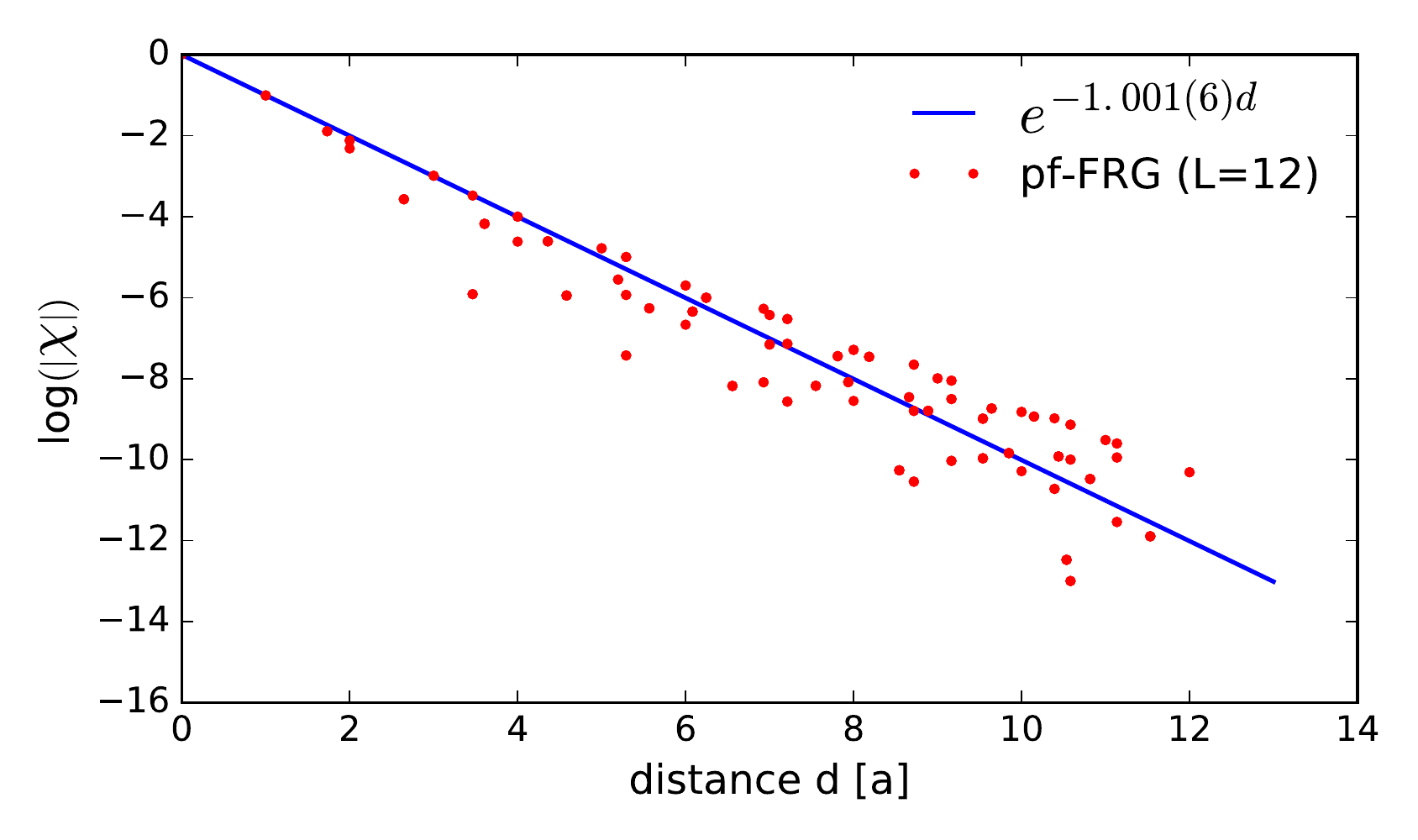}
  \caption{Real space correlations of {\bf spin liquid 1 phase} for the $J_1J_2$-Heisenberg model on the kagome lattice.
  		Top panel: Blue circles represent ferromagnetic correlation and orange circles AFM correlation. 
		The correlation strength is encoded in the circles' magnitudes relative to the reference site indicated in grey.
		Bottom panel: Decay of real space correlations as a function of distance in units of the lattice constant $a$. 
		The pf-FRG result (red) is fitted by an exponential decay. }
  \label{fig:kagome-SL1}
\end{figure}

\begin{figure}
  \centering
  \includegraphics[width=0.65\linewidth]{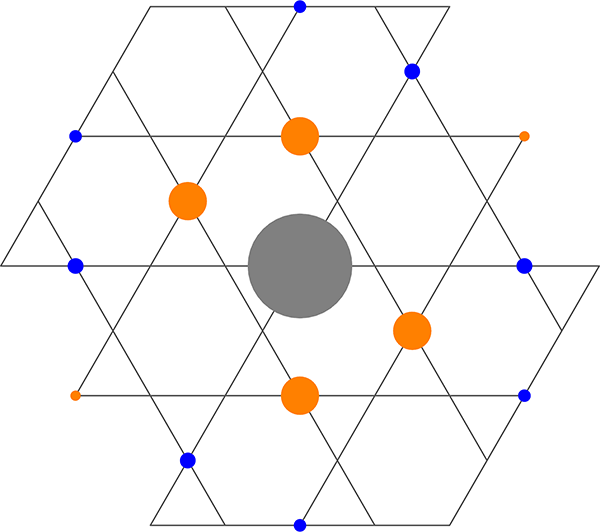} \\
  \includegraphics[width=\linewidth]{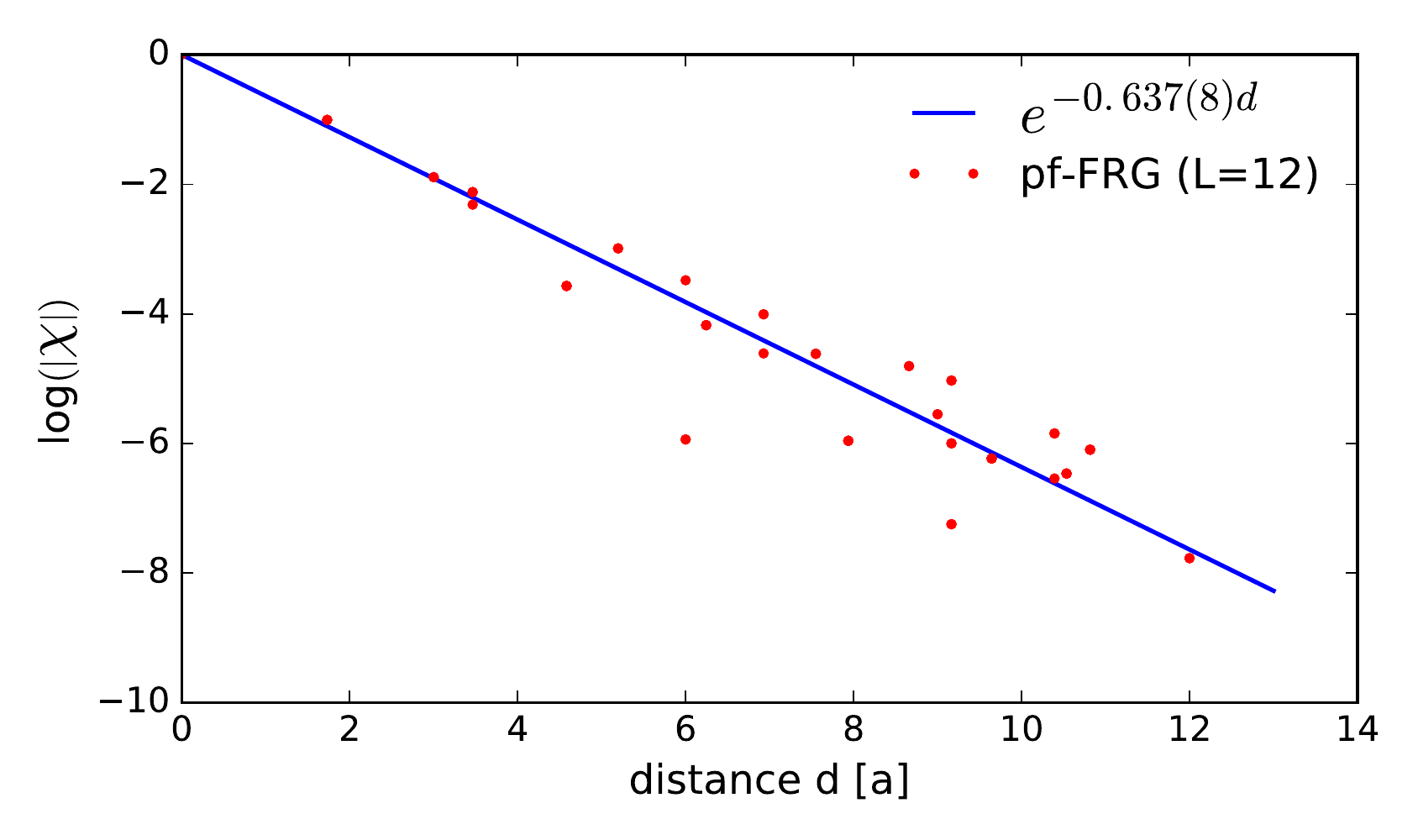}
  \caption{Real space correlations of {\bf spin liquid 2 phase} for the $J_1J_2$-Heisenberg model on the kagome lattice.
  		Top panel: Blue circles represent ferromagnetic correlation and orange circles AFM correlation. 
		The correlation strength is encoded in the circles' magnitudes relative to the reference site indicated in grey. 
		Bottom panel: Decay of real space correlations as a function of distance in units of the lattice constant $a$. 
		The pf-FRG result (red) is fitted by an exponential decay. }
  \label{fig:kagome-SL2}
\end{figure}

The phase diagram of the $J_1J_2$  kagome model further exhibits two spin liquid phases in the vicinity of $\alpha=0$ and $\alpha=\pi/2$, i.e. around the points of purely antiferromagnetic nearest neighbor coupling or next-nearest neighbor coupling only. 
Their origin can be traced back to the geometric frustration induced on the elementary triangles of the kagome lattice by the nearest neighbor coupling $J_1$ and on the two subtriangles of the elementary hexagonal plaquettes induced by the second-nearest neighbor coupling $J_2$, respectively. Note that for the $J_2$ only model the original kagome system decomposes into three sublattices, with each sublattice again forming a kagome lattice.
This also reveals an important structural difference between the kagome and hyperkagome lattices, where for the latter the next-nearest neighbor couplings along angled bonds only was not sufficient to stabilize a spin liquid phase (see also the discussion in the previous Section). For the kagome system both spin liquids are found to be stable over a small sliver of coupling parameters and exhibit rapidly decaying real-space correlations, as illustrated in Figs.~\ref{fig:kagome-SL1} and \ref{fig:kagome-SL2}, with short-distance correlations precisely reflecting the character of the dominant  nearest or next-nearest neighbor exchange, respectively. Such an exponential decay of the spin correlations points to a gapped spin liquid, such as a $Z_2$ spin liquid, which has long been discussed as a possible candidate spin liquid state for the kagome antiferromagnet \cite{Read1991,Wen1991,Sachdev1992,Misguich2002,Wang2006,Lu2011,Depenbrock2012,Jiang2012,Mei2016}.
Despite the  exponential decay of the real-space correlations, the momentum-resolved structure factors of these spin liquid phases shown in Fig.~\ref{fig:j1j2Kagome} (a) and (c) exhibit bow-tie like features (around a pinch point at the $M$-point or $X$-points, respectively), commonly associated with Coulomb phase physics \cite{Henley2010}. Such Coulomb phases arise for frustrated magnets which have local constraints that can be mapped to a divergence-free ``flux" \cite{Henley2010} -- such as pyrochlore antiferromagnets \cite{Isakov2004,Henley2005}, spin-ice systems \cite{Castelnovo2008,Castelnovo2012,Fennell2014}, dimer models \cite{Huse2003,Alet2006,Chen2009} or classical Kitaev models \cite{Chandra2010,Sela2014} -- and reflect the characteristic power-law decay ($1/r^d$ in $d$ spatial dimensions) of the real-space correlations. Here, we do not observe sharp bow-tie features, in accordance with the exponential decay of real-space correlations, which leaves room for multiple interpretations. In the Coulomb gas picture, one could argue that the smeared-out bow-tie features arise from a small violation of the divergence-free flux constraint, i.e. the emergence of a finite monopole density and a subsequent screening giving rise to the exponential correlation decay. An alternative interpretation of the weak bow-tie features is the emergence of a chiral spin liquid \cite{Kalmeyer1987}, some of which have recently been shown to exhibit bow-tie like features in their structure factor \cite{Bieri2015,Bieri2016,Halimeh2016}. Finally, we note that similar bow-tie features are also apparent (but have not been discussed) in recent DMRG calculations \cite{Depenbrock2012,Kolley2015} of the structure factor for small kagome systems in a toroidal geometry as well as variational calculations of projected fermionic wave functions of $U(1)$ spin liquids \cite{Iqbal2013}. On the experimental side, however, inelastic neutron scattering data for herbertsmithite \cite{Han2012,Han2016} and kapellasite \cite{Kapellasite} have seen indications of a small spin gap, but so far not reported the observation of any pinch point-like features in the magnetic structure factor.

\subsection{J$_1$-J$_{2=3}$ model}
\label{sec:kagomeJ1J23}

\begin{figure*}[th!]
  \centering
  \includegraphics[width=\linewidth]{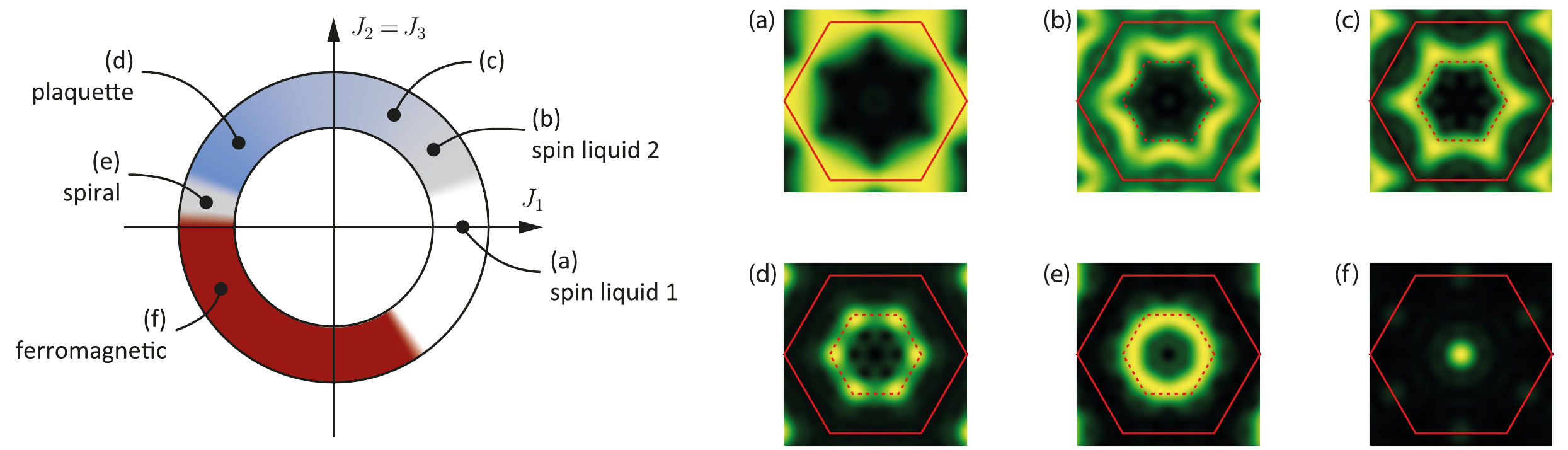}
  \caption{Left panel: Phase diagram of the $J_1J_{2=3}$-Heisenberg model on the kagome lattice for equal strength second- and third-nearest neighbor couplings $J_2=J_3$. 
  	        Right panels: structure factors for the spin liquid and magnetically ordered phases (a)-(f).
	         The extended Brillouin zone is indicated by a solid red line which, if useful, is complemented by the first Brillouin zone 
	         indicated by a dotted line. }
  \label{fig:j1j2bKagome}
\end{figure*}

\begin{figure*}
  \centering
  \includegraphics[width=\linewidth]{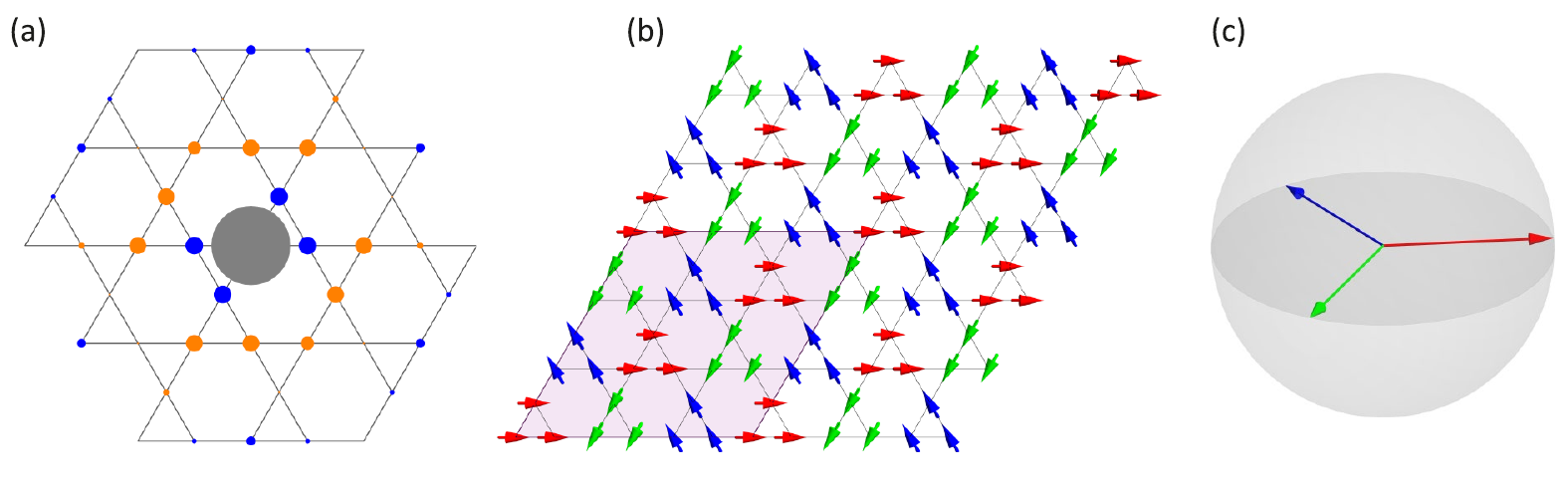}
  \caption{ The {\bf plaquette phase} of the $J_1J_{2=3}$-Heisenberg model on the kagome lattice:
  		a) Real space correlations. Blue circles represent ferromagnetic correlation and orange circles AFM correlation. 
		     The correlation strength is encoded in the circles' magnitudes relative to the reference site indicated in grey.
		b) Spin configuration in real space. 
		c) Relevant axes of magnetic order.
		}
  \label{fig:kagome-plaquette}
\end{figure*}

For the hyperkagome system the third-nearest neighbor coupling along collinear bonds was found to destabilize all non-trivial magnetic orders found in the $J_1J_2$  model, as discussed in detail in Section \ref{sec:hyperkagomeJ1J23}. We expect similar physics to be at play also for the kagome model and therefore investigate the role of the collinear third-nearest neighbor couplings $J_3$ couplings for the two-dimensional situation as well.

\begin{figure}[bh!]
  \centering
  \includegraphics[width=0.9\linewidth]{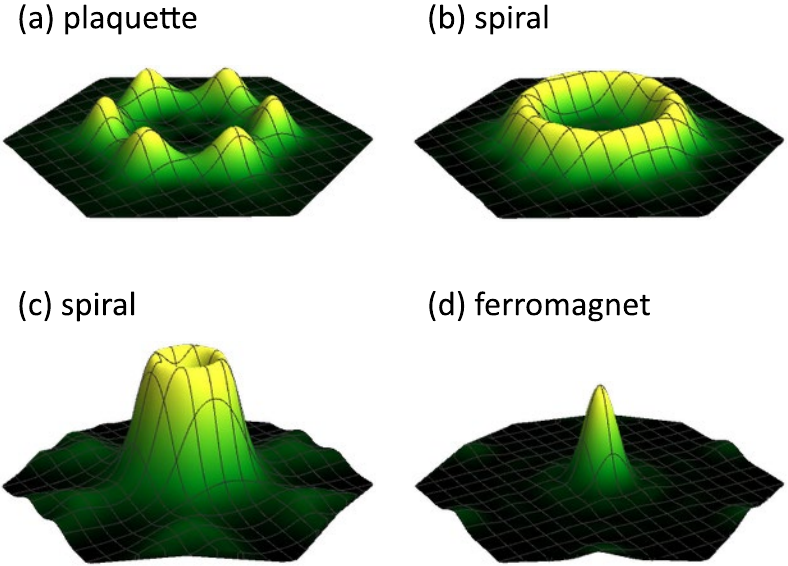}
  \caption{Structure factor evolution around the {\bf spiral phase} of the $J_1J_{2=3}$-Heisenberg model on the kagome lattice. (a) $\alpha=157.5^\circ$, (b) $\alpha=165^\circ$, (c) $\alpha=172.5^\circ$, (d) $\alpha=180^\circ$. 
		  }
  \label{fig:kagome-spiral}
\end{figure}

The phase diagram for equal-strength couplings $J_2 = J_3$ is shown in Fig.~\ref{fig:j1j2bKagome}. In close analogy to the situation for the hyperkagome models, we indeed find that the third-nearest neighbor interaction destabilizes all non-trivial magnetic orders of the $J_1J_2$  model, namely the $\sqrt{3}\times\sqrt{3}$, ${\bf q}=(0,0)$, and cuboc orders. The disappearance of the  $\sqrt{3}\times\sqrt{3}$ and ${\bf q}=(0,0)$ orders can easily be rationalized by the observation that their real-space correlations exhibit opposite correlations for next-nearest and third-nearest neighbors, which are heavily penalized for equal-strength couplings $J_2 = J_3$ in the expanded model. Both phases give way to extended spin liquid regimes, which we will discuss below. For the cuboc order a slightly more subtle picture emerges, since for this type of order both next-nearest and third-nearest neighbors are correlated antiferromagnetically and thus potentially commensurable with the energetics of the expanded model. The cuboc order is nevertheless found to give way to another type of magnetic order upon the inclusion of the third-nearest neighbor interaction. Similar to the hyperkagome system we find the formation of a coplanar plaquette order, in which subsets of spins align ferromagnetically on triangular plaquettes with the plaquettes themselves forming a superlattice $120^\circ$-degree order -- in close analogy to the plaquette order in the hyperkagome system. Further characteristics of this plaquette order are summarized in Fig.~\ref{fig:kagome-plaquette}.

Also the remainder of the  $J_1J_2$ kagome phase diagram (Fig.~\ref{fig:j1j2bKagome}) very closely resembles the phase diagram of the  $J_1J_2$  hyperkagome model (Fig.~\ref{fig:j1j2Hyperkagome}). 
Similar to what we have seen in the three-dimensional model there exists a small, but clearly distinct region between the plaquette ordered phase and the ferromagnetic phase. The structure factor in this region has a characteristic ring-like shape (illustrated in Fig.~\ref{fig:kagome-spiral}) that continuously contracts to just a single peak as one approaches the ferromagnetic phase boundary which might hint at a spiral-like nature of the magnetic order. 
In addition, the two spin liquid regimes already present in the  $J_1J_2$ kagome model are found to be further stabilized by the third-nearest neighbor interaction with their scope in parameter range further expanded, which is again in close analogy to the evolution of the hyperkagome models.


\subsection{J$_1$-J$_2$-J$_3$ model}
\label{sec:hyperkagomeJ1J2J3}

\begin{figure}[t]
  \centering
  \includegraphics[width=\linewidth]{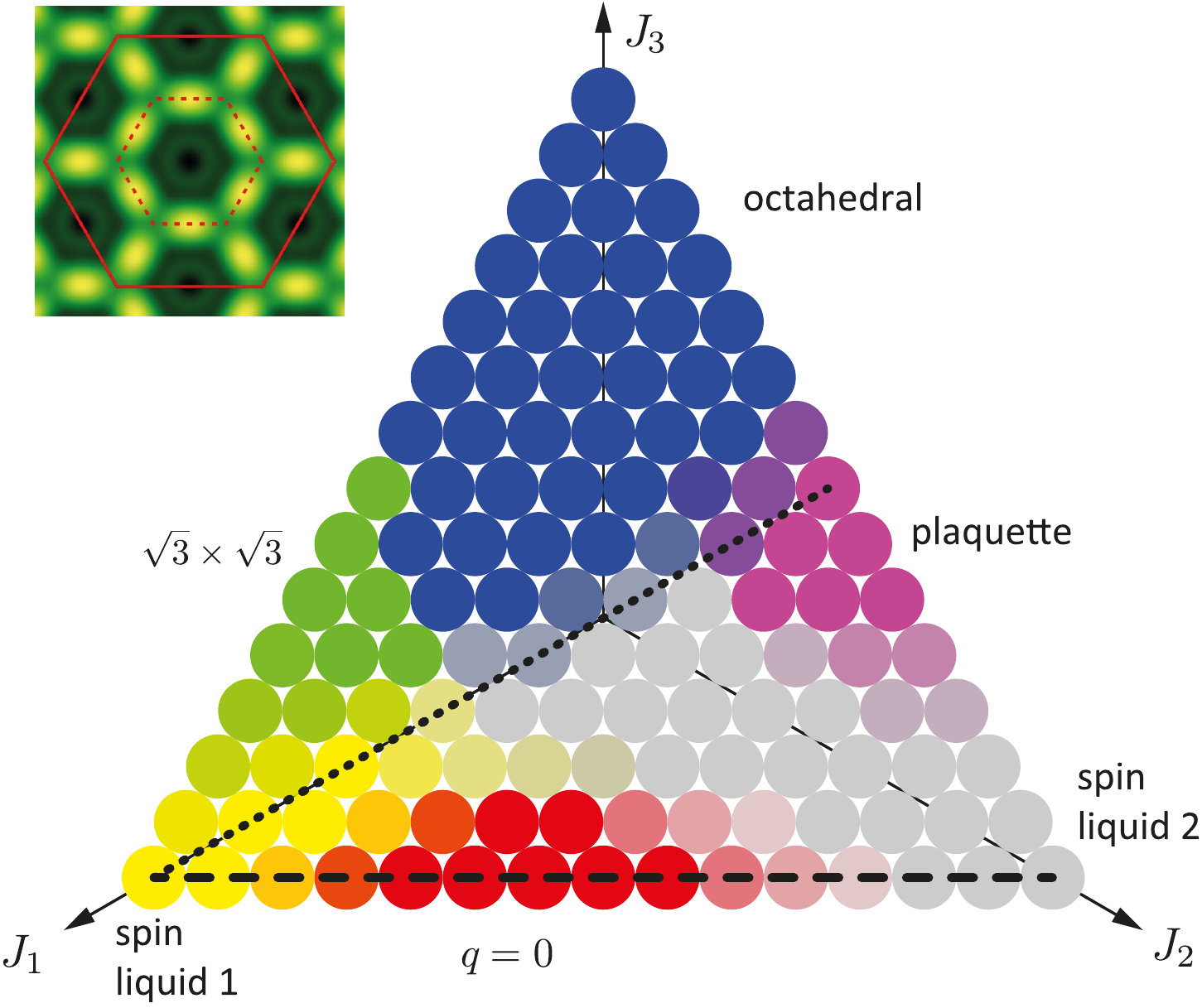}
  \caption{Phase diagram for the $J_1J_2J_3$-Heisenberg model on the kagome lattice in the triangular parameter plane 
  		defined by $J_1+J_2+J_3=1$. The color of the filled circles indicate the various magnetically ordered and spin
		liquid regimes.
		Color gradients are applied to coupling parameters for which the character of the magnetic phase cannot be 
		unambiguously resolved within the pf-FRG scheme. Inset: Structure factor for the octahedral phase. The extended Brillouin zone is indicated by a solid red line, the first Brillouin zone by a dotted line. }
  \label{fig:j1j2j3kagome}
\end{figure}

To round off our discussion of the physics induced by next-nearest neighbor couplings in the kagome system we study a generalized $J_1J_2J_3$  model, in which we also vary the relative strength of the second- and third-nearest neighbor couplings along angled and collinear bonds, respectively. We constrain this discussion to the region that is most relevant for the emergence of spin liquid physics, i.e. a parameter regime for which $J_1$, $J_2$ and $J_3$ are all antiferromagnetic. We investigate this parameter space along the plane defined by $J_1+J_2+J_3=1$. The phase diagram of the $J_1J_2J_3$ kagome model on this so-parametrized 
coupling space is illustrated in Fig.~\ref{fig:j1j2j3kagome}. First of all, note that this phase diagram includes parts of  the previously calculated kagome phase diagrams. The case of $J_3=0$, which is equivalent to the upper right quadrant of the phase diagram of the $J_1J_2$-model in Fig.~\ref{fig:j1j2Kagome}, is indicated by the black dashed line at the bottom of the phase diagram in Fig.~\ref{fig:j1j2j3kagome}. Further, the case of $J_3=J_2$, which matches the upper right quadrant of the $J_1J_{2=3}$ model phase diagram of Fig.~\ref{fig:j1j2bKagome}, is indicated by the diagonal dotted line in Fig.~\ref{fig:j1j2j3kagome}.

\begin{figure*}[th!]
  \centering
  \includegraphics[width=\linewidth]{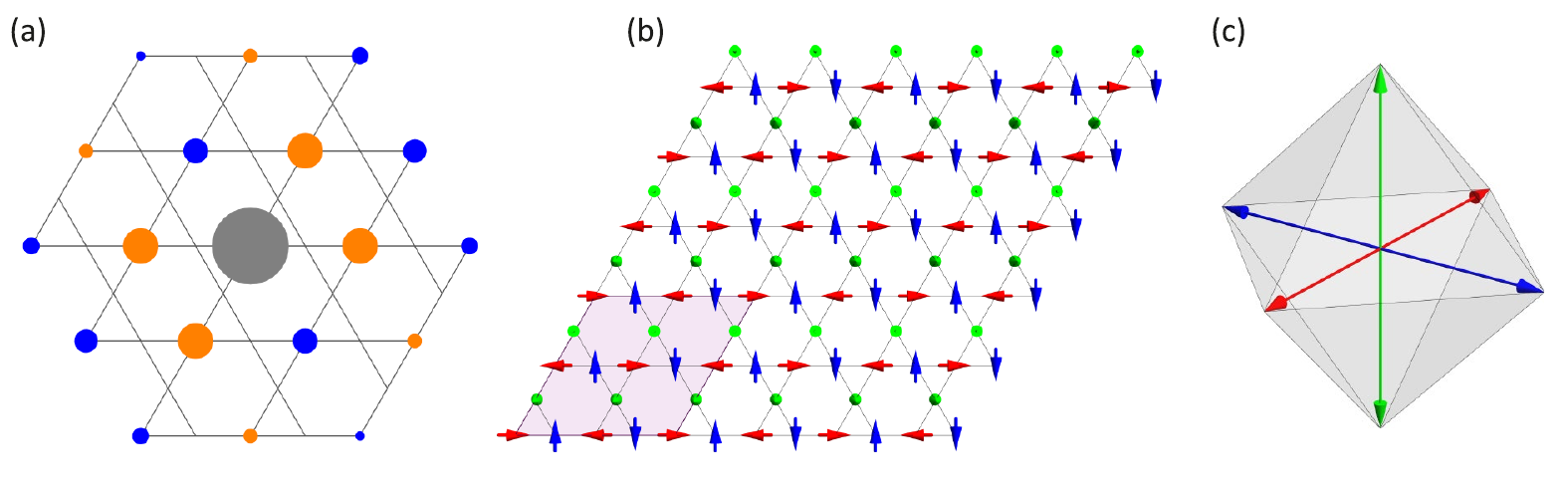}
  \caption{ The {\bf octahedral phase} of the $J_1J_2J_3$-Heisenberg model on the kagome lattice:
  		a) Real space correlations. Blue circles represent ferromagnetic correlation and orange circles AFM correlation. 
		     The correlation strength is encoded in  the circles' magnitudes relative to the reference site indicated in grey.
		b) Spin configuration in real space. 
		c) Relevant axes of magnetic order.
		}
  \label{fig:j1j2j3kagome-octahedral}
\end{figure*}

The phase diagram at hand exhibits four ordered phases with the $\sqrt{3}\times\sqrt{3}$, ${\bf q}=(0,0)$, and plaquette ordered phases now being already familiar acquaintances. The occurrence of a small pocket of $\sqrt{3}\times\sqrt{3}$ order for purely antiferromagnetic couplings is probably noteworthy. While we previously argued that its combination of ferromagnetic next-nearest neighbor correlations with antiferromagnetic third-nearest neighbor correlations (see again the left panel of Fig.~\ref{fig:kagome-sqrt3}) is responsible for the absence of this order in the $J_1J_{2=3}$-model, we here find that antiferromagnetic third-nearest neighbor interactions alone are already sufficient to destabilize the spin liquid of the kagome antiferromagnet ($J_2 = J_3 = 0$) and favor $\sqrt{3}\times\sqrt{3}$ order.
In addition, we find a fourth ordered phase (indicated by the blue dots) in the regime of dominant third-nearest neighbor couplings $J_3$ occupying most of the upper half of the phase diagram. In this phase the kagome lattice decouples into three (deformed) square lattices each of which orders into an antiferromagnetic N\'eel state. Spins of the three different sublattices tend to align perpendicularly and therefore span an octahedron (c.f. Fig. \ref{fig:j1j2j3kagome-octahedral}). This order has been first discussed \cite{Messio2011} in the classical variant of this model and dubbed the octahedral phase. 

Finally, we find two extended spin liquid regimes (indicated by the yellow and grey dots, respectively). From the perspective of this extended phase diagram we are reassured that the spin liquid favored by antiferromagnetic second-nearest neighbor coupling $J_2$ (dubbed ``spin liquid 2" in the phase diagram at hand) is indeed connected to the one that we have previously seen in the $J_1J_{2=3}$ model of Fig.~\ref{fig:j1j2bKagome}.


\section{Discussion}
\label{sec:discussion}

To summarize, we have elucidated the role of next-nearest neighbor couplings on kagome Heisenberg models in two and three spatial dimensions. While the second-nearest neighbor coupling (along angled bonds) stabilizes a variety of magnetic orders for both the kagome and hyperkagome systems, the inclusion of a third-nearest neighbor coupling (along collinear bonds) is found to again destabilize these magnetic orders and in lieu give rise to expanded spin liquid regimes. For the hyperkagome system we have pointed out that -- in contrast to the two-dimensional kagome system -- the nearest-neighbor coupling $J_2$ is in itself not sufficient to induce geometric frustration. Apart from this subtle difference, we generally find rather similar ordering tendencies in the kagome and hyperkagome systems.

On a technical level, we have demonstrated the versatility of the pseudofermion functional renormalization group approach to track the quantum magnetism of frustrated quantum magnets on non-trivial lattice geometries in two and three spatial dimensions. Beyond the numerical efficiency we have demonstrated its quantitative accuracy by direct comparisons to high-temperature series expansion results.
While the pf-FRG approach does extremely well in identifying magnetic orders of both coplanar and non-coplanar character and possibly quite large magnetic unit cells, the identification and characterization of spin liquid physics might leave room for further improvements. It would be highly desirable to expand the current formulation of the pf-FRG approach (i) to incorporate additional terms in the computation of the flow equation that go beyond the Katanin scheme (this has been discussed e.g. in \cite{Eberlein2014, Wentzell2016} in the context of fermionic FRG), (ii) to go beyond SU(2) symmetric Heisenberg models (a first step in this direction was taken for Kitaev-like spin models in Ref.~\onlinecite{Reuther2011c}), (iii) to allow for a positive identification of spin liquids, e.g. via their long-range entanglement, and (iv) to explicitly incorporate gauge structures beyond the $U(1)$ gauge symmetry in order to quantitatively access $Z_2$ spin liquids or chiral spin liquids, which have attracted much recent interest.


\begin{acknowledgments}
We are indebted to J. Reuther for many insightful discussions on the technical aspects of the pf-FRG approach
and thank R.R.P. Singh for a discussion of the high-temperature series expansion results for the hyperkagome lattice \cite{Singh2012}.
This work was partially supported by the DFG within the CRC 1238 (project C03).
The numerical simulations were performed on the CHEOPS cluster at RRZK Cologne.
F.L.B. thanks the Bonn-Cologne Graduate School of Physics and Astronomy (BCGS) for support.
\end{acknowledgments}


\bibliography{kagome}

\begin{thebibliography}{100}
\bibitem [{Foo({\natexlab{a}})}]{FootnoteSpinLiquidCriterion}%
  \BibitemOpen
  \bibinfo {note} {{In the following, we will use precisely this absence of a
  divergence in the magnetic susceptibility (calculated via the pf-FRG scheme)
  down to the lowest temperatures as an indicator of a spin liquid
  phase.}}\BibitemShut {Stop}%
\end{thebibliography}

\end{document}